\documentclass[journal]{IEEEtranTIE}
\usepackage{graphicx}
\usepackage[table]{xcolor}
\usepackage{cite}
\usepackage{picinpar}
\usepackage{amsmath}
\usepackage{url}
\usepackage{flushend}
\usepackage[latin1]{inputenc}
\usepackage{colortbl}
\usepackage{soul}
\usepackage{multirow}
\usepackage{pifont}
\usepackage{color}
\usepackage{alltt}
\usepackage[createShortEnv]{proof-at-the-end}
\usepackage[hidelinks]{hyperref}
\usepackage{enumerate}
\usepackage{siunitx}
\usepackage{epstopdf}
\usepackage{pbox}
\usepackage{esvect}
\usepackage{amssymb}
\usepackage{mathtools}
\usepackage{mathrsfs}
\usepackage{booktabs}
\usepackage{diagbox}
\usepackage{tabularx}
\usepackage{caption}
\usepackage{wrapfig}
\ifCLASSOPTIONcompsoc
\usepackage[caption=false, font=normalsize, labelfont=sf, textfont=sf]{subfig}
\else
\usepackage[caption=false, font=footnotesize]{subfig}
\fi

\newtheorem{remark}{\bf{Remark}}

\begin{document}
\title{Multi-Mode Inverters: A Unified Control Design for Grid-Forming, Grid-Following, and Beyond}

\author{
	\vskip 1em
	
	Alireza Askarian\textsuperscript{1,a}, \emph{Student Member},
	Jaesang Park\textsuperscript{1,b}, \emph{Student Member},
	\\ and Srinivasa Salapaka\textsuperscript{1,c}, \emph{Senior Member}

	\thanks{
	
		\textsuperscript{1} Department of Mechanical Science and Engineering, University of Illinois at Urbana-Champaign, 61801 IL, USA
		
		\textsuperscript{a}askaria2@illinois.edu,\textsuperscript{b}jaesang4@illinois.edu,\textsuperscript{c}salapaka@illinois.edu
		
		This work is supported by the Advanced Research Projects Agency-Energy (ARPA-E) via grant no. DE-AR0001016.
	}
}

\maketitle
	
\begin{abstract}
We present a novel, integrated control framework designed to achieve seamless transitions among a spectrum of inverter operation modes. The operation spectrum includes grid-forming (GFM), grid-following (GFL), static synchronous compensator (STATCOM), energy storage system (ESS), and voltage source inverter (VSI). The proposed control architecture offers guarantees of stability, robustness, and performance regardless of the specific mode. The core concept involves establishing a unified algebraic structure for the feedback control system, where different modes are defined by the magnitude of closed-loop signals. As we demonstrate, this approach results in a two-dimensional continuum of operation modes and enables transition trajectories between operation modes by dynamically adjusting closed-loop variables towards corresponding setpoints. Stability, robustness, and fundamental limitation analyses are provided for the closed-loop system across any mode, as well as during transitions between modes. This design facilitates stable and enhanced on-grid integration, even during GFM operation and weak grid conditions. Ultimately, we demonstrate the key attributes of the proposed framework through simulations and experiments, showcasing its seamless transition in on-grid operation, functionality in islanded mode, and robustness to line impedance uncertainty. 
\end{abstract}

\begin{IEEEkeywords}
Asymmetrical fault, microgrids, grid-forming (GFM) inverter, grid-following (GFL) inverter, virtual inertia, seamless transition, multi-input multi-output (MIMO) systems, phase-locked loop (PLL), robust stability, virtual inertia, weak grid.
\end{IEEEkeywords}

\markboth{IEEE TRANSACTIONS ON INDUSTRIAL ELECTRONICS}%
{}

\definecolor{limegreen}{rgb}{0.2, 0.8, 0.2}
\definecolor{forestgreen}{rgb}{0.13, 0.55, 0.13}
\definecolor{greenhtml}{rgb}{0.0, 0.5, 0.0}

\section{Introduction}

\IEEEPARstart{I}nverters with rapid response times and advanced digital processing capabilities are crucial for large-scale integration of renewable distributed energy resources (DERs) into the power grid, managing significant spatiotemporal variations and inherent uncertainties of renewables. Significant progress has been made in inverter control methods, enabling the integration of renewable energy and dynamic adjustments based on grid conditions, load demand, and energy availability.

However, existing control frameworks for DC/AC inverters are significantly customized and fine-tuned to a {\em specific} and {\em fixed} operating mode based on their intended use, limiting their versatility. For example, in the presence of a stiff grid, inverters operate in grid-following (GFL) mode, where they synchronize and supply power to the grid. On the other hand, in the absence of a viable or stiff grid, inverters usually operate in grid-forming (GFM) mode \cite{rathnayake2021grid}, where they not only control the power flow, but also provide voltage and frequency support at the point of common coupling (PCC). The other popular operating modes are the static synchronous compensator (STATCOM) and energy storage system (ESS), where the inverters serve as ancillary devices (grid support). In STATCOM mode, the inverter improves voltage regulation and power transfer capacity by balancing reactive power. In contrast, an ESS acts as a rapidly dispatchable power source, primarily enhancing frequency control by exchanging active power with the connected AC system.
\begin{figure}[t]
    \centering
    \includegraphics[width = 0.95\columnwidth]{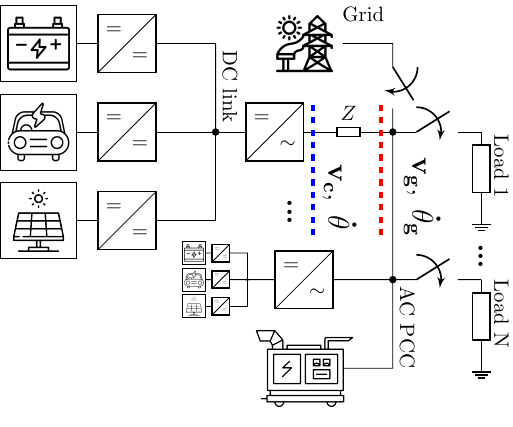}
    \caption{Parallel operation of the collection of inverter units, assimilating distributed energy resource and storage units such as the battery, EV, and renewables into the grid.}
    \label{fig: Parallel_Grid}
    \vspace{0mm}
\end{figure}%
The existing literature focuses predominantly on designing distinct control systems to address the objectives of each mode separately \cite{li2022revisiting}. However, with the increased integration of uncertain DERs and highly dynamic power consumption and generation profiles, it is becoming critical for the same inverter to operate in different operation modes at different times. In this context, the inverter control framework should facilitate a seamless transition between operating modes. For example, when most inverters operate in GFL mode with maximum power point tracking (MPPT), rapid spatiotemporal changes in irradiance (e.g. irradiance anomalies due to moving clouds) lead to rolling and non-localized power imbalance in the network \cite{paletta2023advances}. To address this issue, transitioning some of the inverters near the impacted MPPT areas to ESS or GFM can help locally adjust the power and prevent propagation of undesired voltage and frequency transients in the network.

Another common scenario involves transitioning between grid-tied (GFL) and islanded (GFM) modes, particularly during grid fault. This transition is typically facilitated by reconfiguring and switching the inverter controller based on islanding detection schemes \cite{sadeque2023seamless}\cite{meng2022seamless}. However, most islanding detection methods rely on centralized communication in real time, making them vulnerable to communication delays and blackouts, which limits the robustness of the microgrid \cite{vukojevic2020microgrid}\cite{buduma2022seamless}. Furthermore, sudden switching and reconfiguration of controllers can result in harmful transient response \cite{d2020towards} and instability in low-inertia inverter-based microgrids during the transition \cite{wang2019small}. Consequently, there has been increasing research on achieving a seamless transition in weak grid conditions without control reconfiguration and islanding detection \cite{ashabani2015multivariable}. 

synchronverters, which emulate the behavior of synchronous generators (SG), offer a promising approach to seamless transition \cite{zhong2010synchronverters}\cite{vasudevan2020synchronverter}. These systems replace conventional voltage and current loops with angle, frequency, and torque controllers, providing emulated inertia and enhanced frequency dynamics. However, relying solely on SG dynamics may not fully exploit the fast dynamics of the inverters or allow for more advanced control strategies, potentially resulting in limited stability margins and suboptimal performance. Consequently, numerous studies have focused on improving stability margins \cite{wang2019modified}, damping characteristics \cite{yu2022reference}, and reducing sensitivity to weak grid coupling \cite{roldan2019design} by leveraging inverter dynamics. 

Another approach involves the perpetual operation of the inverters in droop-based grid-forming mode regardless of grid availability \cite{chakraborty2023seamless}\cite{ashabani2015multivariable}. These methods propose dynamically improved droop laws to maintain system stability in both grid-connected and islanded modes. Control strategies based on multivariate loop shaping techniques are explored to decouple active and reactive power in weak grids \cite{ashabani2015multivariable}. Although systematic, these approaches lack multiple-input multiple-output (MIMO) stability analysis, and performance analysis that generalizes beyond the proposed loop shaping controller. \cite{chakraborty2023seamless} provides a mode-dependent droop-based GFM that can operate in a grid-tied mode and improve reactive power regulation upon receiving a mode switch signal. This work provides extensive analysis by modeling the GFM in both on-grid and off-grid operations and investigating the root locus for a range of parameters. However, it does not provide a viable approach to shaping the transient response of the inverter frequency.

This paper introduces a {\em unified} control framework that facilitates seamless transitions between different operating modes. Central to this approach is establishing a common algebraic framework for the feedback control system across various modes. In this structure, the inverter's operational space is viewed as a continuum of modes, including GFM, GFL, STATCOM, and ESS. Each mode is parameterized by two control variables. Smooth transitions between modes are enabled by adjusting these parameters along trajectories in the parameter space, starting and ending at setpoints corresponding to the initial and final modes. The proposed framework transforms all the inverter control objectives into closed-loop properties of the feedback control structure. This approach inherently incorporates the benefits of existing state-of-the-art design concepts such as droop, virtual impedance, PLL, and inertia, without explicitly designing for them. This approach results in a universal characterization of stability, performance, and fundamental trade-offs for the MIMO closed-loop, regardless of the operational mode. The following are the main contributions of this paper:\\
$\bullet$ {\bf Modeling: Universal All-Operating-Mode MIMO Model:} We construct a MIMO model that focuses on the line dynamics that connects the inverter and the grid; also here the inverter is viewed as an actuator.   This viewpoint facilitates control to directly target power variables at the main grid bus avoiding using inverter capacitance voltage as a grid voltage proxy.  
In addition, the entire dynamic system is modeled as a known nominal model plus an unknown perturbation disturbance system, accounting for uncertainties in impedance parameters and AC network influences. Leveraging algebraic structures of known and unknown dynamics, our control design achieves performance goals despite uncertainties.\\
$\bullet$ {\bf Analysis: Robustness and Performance; Fundamental Limitations and Trade-offs; Characterization of all modes} 
We develop a mode-independent analysis framework for the stability and performance of the inverter MIMO closed-loop in terms of a simpler 2-SISO system. Closed-loop sensitivity transfer functions are extensively used to quantify and specify control objectives such as power sharing/tracking, voltage and frequency regulation, synchronization, and inertia despite uncertainties in load and generation profiles. In addition, we quantify the fundamental trade-offs between different control objectives and operation modes using the algebraic properties of the sensitivity transfer function. Finally, we characterize a two-dimensional {\em continuum} of operation modes, including four popular modes (GFL, GFM, STATCOM, ESS), based on the shapes of their sensitivity transfer functions.\\
$\bullet$ {\bf Control Design: Universal Synthesis framework for all modes and inter-mode Transitions:} We propose a control design that balances the desired trade-offs between voltage and frequency regulation, active and reactive power sharing, synchronization and inertia, and robust stability. Furthermore, we show how the proposed controller is implemented using inverter-level closed-loop dynamics. Another important contribution is the seamless transitions between the four fundamental modes (in fact, the continuum of modes). We show that mode transitions can be understood as tracking a trajectory in a two-dimensional parameter space. A Lyapunov-based analysis is proposed to ensure stability during transitions between operating modes.

Finally, through simulations and experiments, we showcase the key attributes of the proposed framework, especially seamless transition between distinct operation modes in both on-grid and islanded mode, power regulation in different scenarios, inertial response, fault ride-through capability, and robustness to line impedance uncertainty. 
\begin{figure}[t]
    \centering
    \includegraphics[width = 0.95\linewidth]{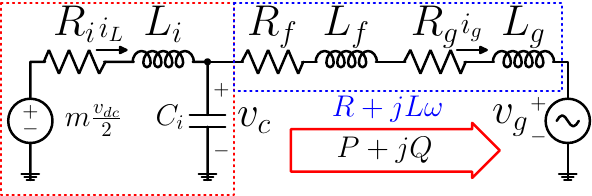}
    \caption{Inverter as a controlled voltage source $v_c$, connected to the Grid $v_g$ via an RL impedance.\label{fig:Averaged_Inverter_and_Line}}
    \vspace{-5mm}
\end{figure}%
\begin{figure}[t]
    \centering
    \subfloat[]
    {
        \includegraphics[width = 0.49\columnwidth]{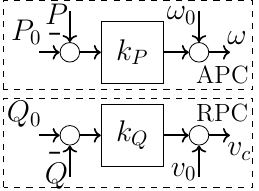}
        \label{fig: APC_RPC}
    }
    \subfloat[]
    {
        \includegraphics[width = 0.49\columnwidth]{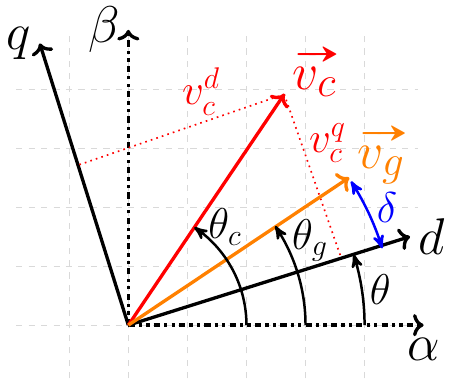}
        \label{fig:DQ_Frame}
    }
    \caption{(a) Active power control (APC) and reactive power control (RPC) based on $P/f$ and $Q/v$ droop. (b) The $dq$ synchronous rotating frame.}
    \vspace{-4mm}
\end{figure}
\section{Problem Setting}
In our framework, we abstract out the inverter interface with a power network, which is common to all operational modes, as two voltage sources connected with an $RL$ impedance as shown in Fig. \ref{fig:Averaged_Inverter_and_Line}. Here, $v_c$ is the inverter's capacitor voltage, while $v_g$ represents the PCC voltage governed by the rest of the microgrid, i.e., the power grid and/or other inverter units. 
The $RL$ impedance signifies the transmission line impedance $R_g+jL_g\omega$ and the grid side impedance of the $LCL$ filter $R_f+jL_f\omega$ ($R_f, L_f=0$ for the $LC$ filter). Regardless of the mode of operation, it is desired that the inverter in Fig. \ref{fig:Averaged_Inverter_and_Line} operate at the fundamental frequency and maintains $v_c$ close to the nominal value of $v_0$ while regulating the active and reactive power flow $\{P,Q\}$ to the desired setpoint $\{P_0,Q_0\}$. However, based on the {\em steady-state} power-flow equation at {\em fundamental frequency} presented below:
% %
{\small\begin{equation}
    \begin{split}
        P &= \left(
        \frac{|v_c||v_g|}{Z}\cos{\delta} -
        \frac{|v_g^2|}{Z}
        \right) \cos{\phi_z}
        -
        \frac{|v_c||v_g|}{Z}
        \sin{\phi_z}\sin{\delta},
        \\
        Q &= \left(
        \frac{|v_c||v_g|}{Z}\cos{\delta} -
        \frac{|v_g^2|}{Z} 
        \right) \sin{\phi_z}
        +
        \frac{|v_c||v_g|}{Z}
        \cos{\phi_z}\sin{\delta},
    \end{split}
    \label{eq:power_flow}
\end{equation}}%
we observe that active and reactive power depend on the inverter and grid voltage magnitudes ($|v_c|$ and $|v_g|$), the phase difference $\delta$ between their voltage phasors (illustrated in Fig. \ref{fig:DQ_Frame}), and the line impedance $Z \angle \phi_z := R + jL\omega$ (shown in Fig. \ref{fig:Averaged_Inverter_and_Line}). Different operating modes, such as the GFM and GFL inverters, differ in the way in which they balance trade-offs between allowable voltage magnitude and frequency deviations in favor of achieving accurate power tracking. For example, the universal droop law in (\ref{eq:universal_droop_law}), or $P/f$ and $Q/v$ active and reactive power control (APC and RPC) in Fig. \ref{fig: APC_RPC}, demonstrate that droop coefficients can be adjusted to manage this trade-off effectively. 
{\small\begin{equation}
    \begin{split}
        \begin{bmatrix}
        |v_c| - v_0 \\ \omega - \omega_0    
        \end{bmatrix}
        =
        \begin{bmatrix}
            k_P & 0 \\
            0 & -k_Q
        \end{bmatrix}
        \begin{bmatrix}
            \hphantom{-}\cos{\phi_z} & \sin{\phi_z} \\
            -\sin{\phi_z} & \cos{\phi_z}
        \end{bmatrix}
        \begin{bmatrix}
            P_0 -P \\
            Q_0 - Q 
        \end{bmatrix}.
    \end{split}
    \label{eq:universal_droop_law}
\end{equation}}%
One of our objectives is to characterize the fundamental trade-offs between control objectives not only in steady state, but throughout the frequency spectrum. Therefore, instead of relying on the {\em nonlinear} {\em steady-state} power flow equation in (\ref{eq:power_flow}) or the dynamic power phasor \cite{sharma2020robust}, we exploit the rich and linear dynamics of the grid current, $i_g$, for the design and analysis of closed-loop control. In this context, the averaged dynamical model of the the transmission line and inverter, as shown in Fig. \ref{fig:Averaged_Inverter_and_Line}, in the direct quadrature ($dq$) frame is
{\small
\begin{align}
        L\frac{d\vv{i_g}}{dt} &=
        \vv{v_c} - \vv{v_g} -
        R\vv{i_g} -
        L\left( \omega_0
        \begin{bmatrix}
        -i_g^q, i_g^d
        \end{bmatrix}^{\top}
        + \psi(t,\vv{i_g})\right),
    \label{eq:line_dynamics_ss_dq} 
    \\
    \begin{split}
    L_i\frac{d\vv{i_L}}{dt} &=
    \frac{v_{dc}}{2}\vv{m} - \vv{v_c} -
    R_i\vv{i_L} -
    L_i\Dot{\theta}
    \begin{bmatrix}
    -i_L^q, i_L^d
    \end{bmatrix}^{\top}, 
    \\
    C_i\frac{d\vv{v_c}}{dt} &=
    \vv{i_L} - \vv{i_g} -
    C_i\Dot{\theta}
    \begin{bmatrix}
    -v_c^q, v_c^d
    \end{bmatrix}^{\top}.
    \label{eq:inverter_dynamics_ss_dq}
    \end{split}
\end{align}}%
Here $\Dot{\theta} = \omega$ is the angular frequency of the $dq$ rotating frame shown in Fig. \ref{fig:DQ_Frame}, $\omega_0$ is the nominal frequency, $\psi(t,\vv{i_g}) = (\Dot{\theta} - \omega_0)\begin{bmatrix} -i_g^q, i_g^d \end{bmatrix}^{\top}$ is sector-bounded nonlinearity, $(v_{dc}/2)\vv{m}$ is the cycle-averaged model of the switching node \cite{6739422}, and each vector contains the respective $d$ and $q$ signal components \cite{6739417}. Applying the Laplace transform to (\ref{eq:line_dynamics_ss_dq}) and (\ref{eq:inverter_dynamics_ss_dq}) we get
\begin{figure}[t]
    \includegraphics[width = \columnwidth]{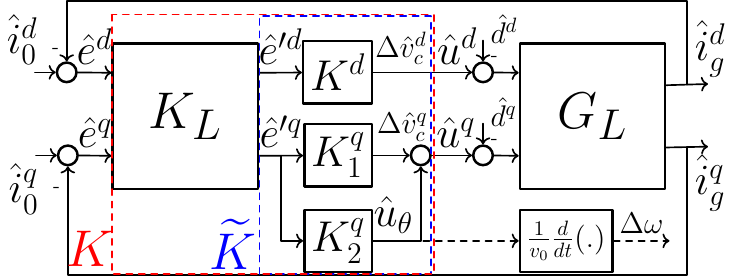}
    
    \caption{
    Proposed closed-loop with line dynamics $G_L$ in (\ref{eq:line_model}) as plant and $K=\widetilde{K}K_L$ as MIMO feedback controller. The $K_L$ is the full matrix part of $K$, while $\widetilde{K}=\text{diag}(K^d,K_1^q + K_2^q)$ is the diagonal part of the controller. The input disturbances $\{d^d, d^q\}$ capture the effect of the PCC on the closed-loop, while $\{u^d,u^q\}$ integrates the capacitor voltage and the $dq$ frame angle $\theta$ as closed-loop control effort.}
    \label{fig: Simple_Feedback}
    \vspace{-3mm}
\end{figure}%
{\small
\begin{align}
    \begin{split}
        \vv{\hat{i}_{g}}
        &=
        \frac{
        \begin{bmatrix}
        s+\lambda & \omega_{0} \\
        -\omega_{0} & s+\lambda
        \end{bmatrix}}
        {L\left( s^2 + 2\lambda s + 
        \lambda^2 + \omega_{0}^{2}\right)}
        \left(\vv{\hat{v}}_c - \vv{\hat{v}}_g\right),
        \,\text{{\normalsize where} $\lambda = \frac{R}{L},$}
    \label{eq:line_model}
    \end{split}\\
    \begin{split}
        \vv{\hat{v}_c} &= 
        \frac{1}{C_i s}
        \left(\vv{\hat{i}_L} - \vv{\hat{i}_g}
        - \mathscr{L} \left\{ 
        C_i \Dot{\theta} \left[-v_c^q,v_c^d\right]^{\top}
        \right\}
        \right),\\
        \vv{\hat{i}_L} &= 
        \frac{1}{L_is+R_i}
        \left(\frac{v_{dc}}{2}\vv{\hat{m}} - \vv{\hat{v}_c}
        - \mathscr{L} \left\{ 
        L_i \Dot{\theta} \left[-i_L^q,i_L^d\right]^{\top}
        \right\}
        \right),
    \label{eq:inverter_ss_dynamics}
    \end{split}
\end{align}}%
where the hat notation signifies the Laplace domain signals, and $\mathscr{L}\{\cdot\}$ is Laplace transform.
\begin{remark}
    The effect of nonlinearity $\psi(\cdot,\cdot)$ is neglected in (\ref{eq:line_model}). However, in Proposition \ref{prop:nonlinear} we establish sufficient conditions for the nominal closed-loop system under which the nonlinearity is sector-bounded and does not compromise closed-loop stability. Therefore, it is justifiable to neglect nonlinearity $\psi$ when the conditions outlined in Proposition \ref{prop:nonlinear} are met \cite{khalil2002nonlinear}.
\end{remark}

Unlike the {\em frequency-invariant} input-output coupling in (\ref{eq:power_flow}), the {\em frequency-dependent} input-output coupling in (\ref{eq:line_model}) allows us to design controllers that achieve improved stability margins by exploiting the dynamic nature of the MIMO coupling.
\section{Closed-Loop Modeling - A New Perspective}
\label{sec: Modelling}
We employ the closed-loop block diagram in Fig. \ref{fig: Simple_Feedback} to analyze the stability and performance of the inverter and grid setup in Fig. \ref{fig:Averaged_Inverter_and_Line}. In this context, $G_L$ represents the MIMO line dynamics in (\ref{eq:line_model}) and the feedback controller $K$, as discussed in Section \ref{sec: mdesign}, encapsulates the inverter dynamics in (\ref{eq:inverter_ss_dynamics}). However, before we evaluate the performance and stability in terms of closed-loop in Fig. \ref{fig: Simple_Feedback}, we need to resolve the control input $\vv{u}=[u^d,u^q]^\top$ and disturbance $\vv{d}=[d^d,d^q]^\top$ in terms of the inverter and grid state variables. 
\begin{propositionE}
\label{prop:control_disturbance}
For the closed-loop structure in Fig. \ref{fig: Simple_Feedback}, the control input $\vv{u}$ and disturbance $\vv{d}$ are 
{\small\begin{equation}
    \begin{split}
    \begin{bmatrix}
    u^d \\ u^q 
    \end{bmatrix}
     &=
    \begin{bmatrix}
    \Delta v_c^d \\ 
    \Delta v_c^q
    \end{bmatrix} + 
    \begin{bmatrix}
    0 \\ u_\theta
    \end{bmatrix}, \; \text{and} \;
    \begin{bmatrix}
    d^d \\ d^q
    \end{bmatrix} =
    \begin{bmatrix}
    \quad\;\;\,\Delta v_g \\
    \|{v_g}\| \int \Delta\omega_g\;dt
    \end{bmatrix},
    \end{split}
    \label{eq:control_disturbance}
\end{equation}}%
where 
{\small\begin{align}
\Delta v_c^d &= v_c^d - v_0, &
\Delta v_c^q &= v_c^q - 0, 
\nonumber
\\
\Delta v_g &= \|v_g\| - v_0, &
\Delta\omega_g &= \omega_g - \omega_0.,
\nonumber
\\
u_\theta(t) &= \|v_g\| \int \Delta \omega\;dt, &
\Delta \omega &= \Dot{\theta} - \omega_0.
\end{align}}%
In the above, $\Delta v_g$ and $\Delta \omega_g$ represent the deviation in the voltage and frequency of the grid from the nominal value.
\end{propositionE}
\begin{proofE}
Based on the closed-loop structure in Fig. \ref{fig: Simple_Feedback} and line dynamics in (\ref{eq:line_model}) we have $\vv{u} - \vv{d} = \vv{v}_c - \vv{v}_g = (\vv{v}_c - [v_0,0]^\top) - (\vv{v}_g - [v_0,0]^\top)$. Therefore, following relationship holds
{\small\begin{align}
    \vv{u} - \vv{d}
    &=
    \begin{bmatrix}
        \Delta v_c^d \\ 
        \Delta v_c^q
    \end{bmatrix}
    -
    \begin{bmatrix}
    \|{v_g}\|\cos(\delta) - v_0
    \\
    \|{v_g}\|\sin(\delta) \hphantom{\;\, - v_0}
    \end{bmatrix} 
    \nonumber
    \\
    & =
    \begin{bmatrix}
        \Delta v_c^d \\ 
        \Delta v_c^q
    \end{bmatrix}
    -
    \begin{bmatrix}
    \Delta v_g
    \\
    \|{v_g}\|\delta
    \end{bmatrix} + \mathcal{O}\left(\delta^2\right).
    \label{eq:Control_Disturbance_Proof}
\end{align}}%
In above, $\delta=\theta_g - \theta$ is the angle between the grid voltage phasor $\vv{v_g}$ and the $dq$ rotating frame as shown in Fig. \ref{fig:DQ_Frame}. Moreover, assuming small-angle condition, that is $\mathcal{O}\left(\delta^2\right)\approx 0$ in (\ref{eq:Control_Disturbance_Proof}), we can write $\|v_g\|\delta = d^q - u_\theta$ where
{\small\begin{align}
    d^q &= 
    \|{v_g}\| \int_{t_0}^{t} (\Dot{\theta}_g(\tau) - \omega_0)\;d\tau,
    &
    u_\theta &=
    \|{v_g}\| \int_{t_0}^{t} (\Dot{\theta}(\tau) - \omega_0)\;d\tau.
    \label{eq:vq_disturbance_control}
    \nonumber
\end{align}}%
Here, $\Dot{\theta}_g$ is grid frequency and $\Dot{\theta}$ is the $dq$ frame angular frequency. Ultimately, capacitor voltage, $\Delta\vv{v_c}$, and $dq$ rotating angle, $\theta$ and subsequently $u_\theta$, are controllable variables, while the grid states, $\Delta v_g$ and $\theta_g$, are not measurable or even controllable. Therefore, we recover (\ref{eq:control_disturbance}).
\end{proofE}

\begin{remark}
Grid disturbance $\vv{d}$ is unknown, however we adopt following disturbance model for closed-loop analysis
{\small\begin{equation}
    \begin{split}
        \begin{bmatrix}
            \hat{d}^d \\ \hat{d}^q
        \end{bmatrix} 
        =
        \begin{bmatrix}
            1 & 0\\
            0 & v_0/s
        \end{bmatrix}
        \begin{bmatrix}
            \Delta v_g/s \\
            \Delta\omega_g/s
        \end{bmatrix}
        +
        \vv{\hat{\eta}}(s).
    \end{split}
    \label{eq:disturbance_type}
\end{equation}}%
Above disturbance model assumes nominally constant deviations in grid's voltage and frequency, with $\vv{\hat{\eta}}(s)$ representing unmodeled disturbances. We can update the model in (\ref{eq:disturbance_type}) if additional information on grid dynamics is available. For example, in the case of a line-to-ground fault, we can include the second-order harmonic model in (\ref{eq:disturbance_type}) \cite{askarian2024enhanced}.
\end{remark}
Closed-loop in Fig. \ref{fig: Simple_Feedback} maps the current setpoint $\vv{i_0}$, and grid disturbance $\vv{d}$ into the grid current $\vv{i_g}$ and control signal $\vv{u}$ as %
{\small\begin{align}
    \vv{\hat{i}_g} &= G_L S (K \vv{\hat{i}_0} - \vv{\hat{d}}),
    &
    \vv{\hat{u}} &= T(G_L^{\text{-}1} \vv{\hat{i}_0} + \vv{\hat{d}}),
    \label{eq:Grid_Current_Tracking}
\end{align}}%
where the closed-loop sensitivity $S$ and complementary sensitivity $T$ transfer functions are defined as
{\small\begin{align}
    S &= \left(I_2 + K G_L\right)^{\text{-}1},
    &
    T &= I_2 - S
    = KG_L\left(I_2 + K G_L\right)^{\text{-}1}.
    \label{eq: sensitivity}
\end{align}}%%
It is clear from (\ref{eq:Grid_Current_Tracking}) that the closed-loop objectives and internal stability hinge on the \emph{sensitivity} $S$, and \emph{complementary sensitivity} $T$ transfer functions in (\ref{eq: sensitivity}) \cite{skogestad2005multivariable}. Furthermore, both $S$ and $T$ are shaped by designing the feedback controller $K$.
\subsection{Transforming MIMO Closed-Loop System into a SISO System with Coupled Perturbation}
The plant $G_L$, the feedback compensator $K$, the sensitivity $S$ and the complementary sensitivity $T$ in (\ref{eq:Grid_Current_Tracking}) and (\ref{eq: sensitivity}) are coupled MIMO systems. The complexity of designing, evaluating the stability, and analyzing the performance of MIMO closed-loop control is significantly greater than that of SISO closed-loop control. In this work, we mitigate the MIMO coupling problem by factoring closed-loop sensitivity $S$ into diagonal (2-SISO) sensitivity $\widetilde{S}$ and a multiplicative perturbation $\Gamma\mathcal{X}_c$; that is, $S=\Gamma\mathcal{X}_c\widetilde{S}$. Subsequently, we derive the bounds for the MIMO perturbation $\|\Gamma\mathcal{X}_c\|$ and convert the MIMO analysis and synthesis problems into a simpler SISO problem based on the nominal 2-SISO sensitivity $\widetilde{S}$.

We obtain this factorization by first decomposing the controller $K$ as follow 
{\small\begin{align}
    K &= \widetilde{K}K_L,
    &&\text{\normalsize where}&
    \widetilde{K} &= \text{diag}
    (K^d,K^q),
    \;
    K_L \in 
    RH_\infty.
    \label{eq:composit_control}
\end{align}}%
Essentially, $\widetilde{K}$ is diagonal controller, and $K_L$ is a proper and stable full transfer matrix. Also, in the context of the plant $G_L$ in Fig. \ref{fig: Simple_Feedback}, we define the modified plant $G_M$ and its corresponding diagonal part $\widetilde{G}_M$ as
{\small\begin{align}
    G_M &= K_L G_L,
    &
    \widetilde{G}_M &= \text{diag}(G_{M}^d,G_{M}^q),
    \label{eq:aug_plant}
\end{align}}%
where $G_M^{d}$ and $G_M^q$ are scalar transfer functions that represent first and second diagonal elements of $G_M$. We can write $G_M$ as a perturbed version of the diagonal plant $\widetilde{G}_M$ as follows
{\small\begin{align}
    G_M &= \widetilde{G}_M\left(I_2 + E\right), 
    &
    \;\text{\normalsize where}&\;
    &
    E &= \widetilde{G}_M^{\text{-}1}( G_M - \widetilde{G}_M).
    \label{eq: diag_inter}
\end{align}}%
Finally, based on (\ref{eq:composit_control}), (\ref{eq:aug_plant}), and (\ref{eq: diag_inter}) we derive the 2-SISO sensitivity $\widetilde{S}$, and its perturbation $\Gamma\mathcal{X}_c$. 
\begin{propositionE}
\label{prop:SISO_Factorization}
(a) \emph{Factorization of Sensitivity:} We factor the MIMO sensitivity in (\ref{eq: sensitivity}) into a nominal 2-SISO sensitivity $\widetilde{S}=\text{diag}(\widetilde{S}^d,\widetilde{S}^q)$ and coupled perturbation $\Gamma\mathcal{X}_c$ as
{\small
\begin{align}
S &=
\Gamma \mathcal{X}_c \widetilde{S},
\;\text{{\normalsize where}}\;
\Gamma = G_M^{\text{-}1} \widetilde{G}_M,
\label{eq:sensitivity_gamma}
\\
\mathcal{X}_c &= (I_2 + \widetilde{S} (\Gamma - I_2) )^{\text{-}1},
\;\text{{\normalsize and}}
\label{eq:Cross_Coupling_Term}
\\
\widetilde{S}
&=
(I_2 + \widetilde{K}\widetilde{G}_M )^{\text{-}1}
=
\begin{bmatrix}
(1 + K^d G_M^d)^{\text{-}1} & 0 \\
0 & (1 + K^q G_M^q)^{\text{-}1}
\end{bmatrix}.
\label{eq:SISO_Sensitivity}
\end{align}}
(b) \emph{Coupling Strength:} We can bound the coupling effect of $\mathcal{X}_c$ using 
$\epsilon(\omega):=
    \|\widetilde{S}\left(j\omega\right)\|_2 
    \|\Gamma\left(j\omega\right) - I_2\|_{2}
$ as follow
{\small\begin{align}
    &\|\mathcal{X}_c(j\omega) - I_2\|_2
    \leq
    \frac{\epsilon\left(\omega\right)}{1 - \epsilon\left(\omega\right)}.
    \label{eq:magnitude_condition}  
\end{align}}%
(c) \emph{Closed-Loop Characterization:}  Effect of the coupling $\mathcal{X}_c$ and 2-SISO sensitivity $\widetilde{S}$ on closed-loop dynamics in (\ref{eq:Grid_Current_Tracking}) is
{\small\begin{align}
\vv{\hat{i}_g'} = 
\widetilde{G}_M
\mathcal{X}_c\widetilde{S}
\left(\widetilde{K}\vv{\hat{i}_0} - \vv{\hat{d}}\right),
\;
\vv{\hat{u}} = 
\widetilde{K}\widetilde{G}_M
\mathcal{X}_c\widetilde{S}
\left(G_M^{\text{-}1}\vv{\hat{i}_0} + 
\vv{\hat{d}}\right).
\label{eq: Control_Effort_Coupled}
\end{align}}%
\end{propositionE}
\begin{proofE}
(a) Using $K = \widetilde{K}K_L$ and $G_M = K_L G_L$ we get 
{\small\begin{equation}
    \begin{split}
        S
        =
        ( I_2 + K G_L )^{\text{-}1}
        =
        ( I_2 + \widetilde{K}K_L G_L )^{\text{-}1}
        =
        (I_2 + \widetilde{K}G_M)^{\text{-}1}.
    \end{split}
\end{equation}}%
Replacing the modified plant $G_M$ by $\widetilde{G}_M(I_2 + E)$ we get
{\small
\begin{align}
    \begin{split}
        (I_2 + \widetilde{K}\widetilde{G}_M
        (I_2 + E))^{\text{-}1}
        =
        (I_2 + \widetilde{S}\widetilde{K}
        \widetilde{G}_ME )^{\text{-}1}
        (I_2 + \widetilde{K}\widetilde{G}_M )^{\text{-}1},
        \nonumber
    \end{split}
\end{align}}%
We use the identity $\widetilde{S}\widetilde{K}\widetilde{G}_M = I_2 - \widetilde{S}$ in above to get
{\small\begin{equation}
    \begin{split}
        S = 
        (I_{2} + (I_{2} - \widetilde{S})E
        )^{\text{-}1}
        (I_{2} + \widetilde{K}\widetilde{G}_M
        )^{\text{-}1}.
        \label{eq:sensitivity_factorization}
    \end{split}
\end{equation}}%
As final step, we replace $E$ by $\left(\Gamma^{\text{-1}} - I_2\right)$ and factor out $\Gamma^{\text{-}1}$ to arrive at the sensitivity factorization in (\ref{eq:sensitivity_gamma}). $\blacksquare$\\
(b) We use the Neumann series to get
{\small\begin{align}
    \mathcal{X}_c - I_2= 
    (I_2 + \widetilde{S}(\Gamma - I_2)
    )^{\text{-}1} - I_2
    =
    \sum_{n=1}^\infty (-\widetilde{S}(
    \Gamma - I_2
    ))^n.
    \nonumber
\end{align}}%
Right-hand side of above is bounded by geometric series below
{\small\begin{align}
    \begin{split}
        \sum_{n=1}^\infty \left(\|
        \Gamma\left(j\omega\right) - I_2
        \| 
        \|\widetilde{S}\left(j\omega\right)\|_2\right)^n
        = \frac{\epsilon\left(\omega\right)}{1 - \epsilon\left(\omega\right)}
        .\; \blacksquare
    \end{split}
    \nonumber
\end{align}}
(c) The proof follows from replacing the sensitivity factorization (\ref{eq:sensitivity_gamma}) into (\ref{eq:Grid_Current_Tracking}) and (\ref{eq: sensitivity}).
\end{proofE}

\begin{remark}
Based on (\ref{eq:magnitude_condition}), we can design for small $\epsilon$ by (a) reducing sensitivity $\widetilde{S}$ through a control design for $\widetilde{K}$ or (b) shaping $\Gamma$ close to identity. In this work, we assume that the design satisfies $\|\epsilon\|_\infty \leq 0.1$ and that the approximation $\mathcal{X}_c\approx I_2$ is accurate for transient and steady-state analysis. Therefore, we simplify (\ref{eq: Control_Effort_Coupled}) into
{\small\begin{align}
\begin{split}
    \vv{\hat{i}_g'} &= 
    \widetilde{T} \vv{\hat{i}_0'} 
    - \widetilde{G}_M\widetilde{S}\vv{\hat{d}},
    \\
    \vv{\hat{u}} &= 
    \widetilde{T}
    (G_M^{\text{-}1} \vv{\hat{i}_0'} + 
    \vv{\hat{d}}),
    \\
    \vv{\hat{e}'} &=
    \widetilde{S}
    (\vv{\hat{i}_0'} + \widetilde{G}_M\vv{\hat{d}}),
\end{split}
\label{eq:SISO_Effort_Simplified}
\end{align}}%
where 
{\small\begin{align}
\vv{i'_g} &= K_L \vv{i_g}, &
\vv{i'_0} &= K_L \vv{i_0}, &
\vv{e}' &= K_L \vv{e},
\end{align}}
and $\widetilde{T}$ is a 2-SISO complementary sensitivity defined below
{\small\begin{align}
    \widetilde{T} 
    =
    I_2 - \widetilde{S}
    =
    \widetilde{S}\widetilde{K}\widetilde{G}_M
    =
    \begin{bmatrix}
        \widetilde{S}^d K^d G_M^d & 0
        \\
        0 & \widetilde{S}^q K^q G_M^q
    \end{bmatrix},
    \label{eq:SISO_Complementary_Sensitivity}
\end{align}}% 
\end{remark}
The primary transfer matrices in (\ref{eq:SISO_Effort_Simplified}) are diagonal as denoted by the $\sim$ notation. This allows us to circumvent the difficulties of MIMO control and develop a simplified but comprehensive framework for the design and analysis of closed-loop control.
\section{Characterization of MIMO Stability In Terms of SISO and Coupling Perturbation Stability}       
\label{sec: SISO_Stability}
We exploit the SISO factorization of $S$ in Proposition \ref{prop:SISO_Factorization} to propose novel stability criteria for control synthesis.
\begin{propositionE}
\label{prop:MIMO_Stability}
The internal stability of the closed-loop in Fig. \ref{fig: Simple_Feedback} is guaranteed if (a) holds with either (b) or (c).\\
(a) \emph{SISO Condition:} $\widetilde{S}^d$, and $\widetilde{S}^q$ in (\ref{eq:SISO_Sensitivity}) are stable.\\
(b) \emph{Decoupling Condition:} $G_M$ is triangular or diagonal.\\
(c) \emph{Magnitude Condition:}
$\epsilon$ in (\ref{eq:magnitude_condition})
satisfies $\|\epsilon\left(\omega\right)\|_{\infty}< 1$.
\end{propositionE}
\begin{proofE}
The plant $G_L$ and the controller $K$ are stable and minimum phase (we consider stable and minimum phase controllers); hence, the closed-loop in Fig. \ref{fig: Simple_Feedback} is internally stable if and only if $S$ is stable \cite{skogestad2005multivariable}. The stability of $\widetilde{S}$ and $\mathcal{X}_c$ is sufficient for the stability of $S$ ($\Gamma$ is stable if $K_L$ is minimum phase). Condition (a) assumes that $\widetilde{S}$ is stable; The stability of the coupling term $\mathcal{X}_c$ is guaranteed through the spectral radius condition
$
    \rho( \widetilde{S}(j\omega)
    ( \Gamma\left(j\omega\right)-I_{2} ) )
    < 1,\; \forall\omega
$. The $\epsilon$ as defined in (\ref{eq:magnitude_condition}) is an upper bound on the spectral radius; therefore, the magnitude condition in (c) provides a sufficient but conservative guarantee for the stability of $\mathcal{X}_c$. The condition in (b) is less conservative than the condition in (c) and directly guarantees that the spectral radius is zero, regardless of the magnitude and shape of $\widetilde{S}$. If $G_M$ is triangular, then $\Gamma - I_2 = G_M^{\text{-}1}\widetilde{G}_M - I_2$ is strictly triangular and, consequently, $\widetilde{S}\left(\Gamma - I_2\right)$ is also strictly triangular. All the eigenvalues of a strictly triangular matrix are zero; hence, irrespective of $\widetilde{S}$, we have a spectral radius equal to zero.$\blacksquare$
\end{proofE}

From this theorem, we can guarantee stability of $\mathcal{X}_c$ by either reducing the magnitude of $\widetilde{S}$ or shaping the modified plant $G_M$ into a triangular or diagonal matrix. Maintaining two distinct methods is essential, since reducing $\widetilde{S}$ requires a high-gain controller $\widetilde{K}$ (see (\ref{eq:SISO_Sensitivity}) which is not always feasible. For instance, in GFM operation, as we shall see later, the low-frequency gain of the controller is bounded above, limiting the feasible reduction in sensitivity at low frequencies. Moreover, in any case, $\widetilde{S}$ converges to the identity at high frequencies for a proper feedback controller. 

\subsection{The Algebraic Structure of Plant Shaping Controller}
In scenarios where it is not possible to arbitrarily reduce $\widetilde{S}$ (e.g., GFM operation), Proposition \ref{prop:MIMO_Stability} (b) imposes a structure on the transfer matrix $K_L$ such that the modified plant $G_M=K_LG_L$ is triangular or diagonal. To characterize such a structure, we consider a static row normalized $K_L$ as      
{\small\begin{equation}
    K_L = 
    \begin{bmatrix}
        \cos{\phi_1} & -\sin{\phi_1}
        \\
        \sin{\phi_2} & \hphantom{-}\cos{\phi_2}
    \end{bmatrix},
    \label{eq:General_KL}
\end{equation}}%
where $\phi_1,\phi_2\in \left[0,\pi/2\right]$. In the following proposition, we associate the parameters $\phi_1$ and $\phi_2$ with the stability margin of the coupling term $\mathcal{X}_c$ and the conditioned number of the modified plant $G_M = K_L G_L$. Subsequently, we present a transfer function $K_L(s)$ that simultaneously achieves optimal stability margins at both low and high frequencies.
\begin{propositionE}
    \label{prop: rotation_stability}
    (a) The modified plant $G_M(j0)$ satisfies Proposition \ref{prop:MIMO_Stability} (b) in steady state if $\phi_1=\phi_z$ or $\phi_2 = \phi_z$. Here, $\phi_z = \arctan X/R$ where $X=L\omega_0$ and $R$ are the transmission line reactance and resistance.\\
    (b) At high frequency ($s\to\infty$), the stability margin of the coupling term $\mathcal{X}_c$ for $K_L$ in (\ref{eq:General_KL}) converges to
    {\small\begin{equation}
        \lim_{s \to \infty}\det{\left(\mathcal{X}_c\right)} =
        \frac{\cos{\left(\phi_1 - \phi_2\right)}}
        {\cos{\phi_1}\cos{\phi_2}},
    \end{equation}}%
    For $\phi_1=0$ or $\phi_2=0$, the stability margin is 1 (optimal) irrespective of the line impedance phase angle $\phi_z$. Hence, a static $\mathcal{X}_c$ can achieve an optimal stability margin at low and high frequencies if $\{\phi_1=\phi_z,\phi_2=0\}$ or $\{\phi_1=0,\phi_2=\phi_z\}$.
    \\
    (c) The condition number of the modified plant $G_M = K_LG_L$ at both low and high frequencies is given as
    {\small\begin{equation}
        \lim_{\omega \to \infty}
        \kappa\left(G_M(j\omega)\right) =
        \kappa\left(G_M(j0)\right) = 
        \sqrt{\frac{1 + \|\sin{(\Delta \phi)}\|}
        {1 - \|\sin{(\Delta \phi)}\|}},
        \label{eq: Cond_Number}
    \end{equation}}%
    where $\Delta\phi = \phi_1 - \phi_2$. A small condition number indicates the robustness of the plant to input uncertainty. A large condition number, however, implies lack of effective control over plant output associated with smallest singular value. We achieve the smallest possible condition number of one for $\phi_1=\phi_2$. Static $K_L$ cannot satisfy (a) and (b) and achieve the optimal condition number simultaneously, except for resistive lines.
\end{propositionE}
\begin{proofE}
    (a) define $Z = \sqrt{X^2 + R^2}$ and $\phi_1 = \phi_z$ then
    {\small\begin{equation}
        G_M\left(j0\right) 
        =
        K_L\left(j0\right)G_L\left(j0\right)
        =
        \frac{1}{Z}
        \begin{bmatrix}
        1 & 0
        \\
        \sin{\left(\phi_2 - \phi_z\right)} & 1
        \end{bmatrix}.
    \end{equation}}%
    The case for $\phi_2 = \phi_z$ follows the same but instead results in an upper triangular matrix.\\
    (b) we have
    {\small\begin{align}
        \lim_{s \to \infty}\widetilde{S} &= I_2,
        &
        &\text{therefore}
        &
        \lim_{s \to \infty}\text{det}(\mathcal{X}_c) &= \lim_{s \to \infty}\text{det}(\Gamma)^{\text{-}1}.
        \nonumber
    \end{align}}%
    Moreover, $\text{det}(\Gamma)^{\text{-}1} = \text{det}(K_L)\text{det}(G_L)/\text{det}(\widetilde{G}_m)$. Taking into account that $\text{det}(K_L)=\cos{(\phi_1 - \phi_2)}$ we get  
    {\small\begin{equation}
        \lim_{s \to \infty} \text{det}(\Gamma)^{\text{-}1}
        =
        \cos{(\phi_1 - \phi_2)}
        \lim_{s \to \infty} 
        \frac{\text{det}(G_L)}{\text{det}(\widetilde{G}_M)}
        =
        \frac{\cos{(\phi_1 - \phi_2)}}{\cos{(\phi_1)}\cos{(\phi_2)}},
        \nonumber
    \end{equation}}
    and the proof is complete. $\blacksquare$
    \\
    (c) For an arbitrary $\phi_1,\phi_2 \in \left[0,\pi/2\right]$ we have 
    {\small\begin{equation}
        G_M\left(j0\right) =
        \frac{1}{Z}
        \begin{bmatrix}
            \cos{\left(\phi_1 - \phi_z\right)} &
            -\sin{\left(\phi_1 - \phi_z\right)} \\
            \sin{\left(\phi_2 - \phi_z\right)} &
            \hphantom{-}\cos{\left(\phi_2 - \phi_z\right)}
        \end{bmatrix},
    \end{equation}}%
    and therefore $\Bar{\sigma}\left(G_M\left(j0\right)\right) = \sqrt{1 + \|\sin{\left(\phi_1 - \phi_2\right)}\|}$, $\underline{\sigma}\left(G_M\left(j0\right)\right) = \sqrt{1 - \|\sin{\left(\phi_1 - \phi_2\right)}\|}$. $\blacksquare$
\end{proofE}

The Proposition \ref{prop: rotation_stability} outlines a design for a static $K_L$. We can use the same analysis to obtain the transfer matrix $K_L(s)$. One particular choice that satisfies the stability conditions in (a), (b), and the optimal condition number (\ref{eq: Cond_Number}) is
{\small\begin{align}
K_L(s) = 
&\frac{\omega_m}{s + \omega_m}
\begin{bmatrix}
    \frac{L}{Z}s + \cos{\phi_z} & 
    -\sin{\phi_z}
    \\
    \sin{\phi_z} &
    \frac{L}{Z}s + \cos{\phi_z}
\end{bmatrix}.
\label{eq:KL_Diagonal}
\end{align}}%
This of $K_L$  result in the diagonal modified plant $G_M$ below
{\small\begin{align}
    G_M = K_L G_L =  
    \frac{\omega_m/Z}{s + \omega_m}I_2
    =
    \widetilde{G}_M.
    \label{eq:GM_Modified}
\end{align}}%
\begin{remark} Many choices for the transfer matrix $K_L(s)$ satisfy the stability conditions in Proposition \ref{prop: rotation_stability} (a) and (b). The proposed $K_L$ in (\ref{eq:KL_Diagonal}) is one such a choice that also achieves an optimal condition number. As we show in Remark \ref{rem:exact_sharing}, we can adopt a $K_L(s)$ with a suboptimal condition number that offers more attractive power sharing performance. 
\end{remark}
\subsection{Stability Analysis of Sector-Bounded Nonlinearity}
When deriving the linear Laplace model in (\ref{eq:line_model}) from the nonlinear line dynamics in (\ref{eq:line_dynamics_ss_dq}), we neglected the nonlinear term $\psi(\cdot,\cdot)$. To justify this omission and validate the linear stability analysis presented in this section, it is necessary to establish conditions that ensure the closed-loop stability in the presence of this nonlinearity. Assuming the nonlinear term is sector-bounded, we derive such a condition condition-expressed in terms of closed-loop sensitivity-that guarantees stability. This condition is presented in the following proposition, and its implications for feedback control design are discussed in Section \ref{sec: Nonlinear_PR_Design}.
\begin{propositionE}
\label{prop:nonlinear}
    The closed-loop system in Fig. \ref{fig: Simple_Feedback} maintains stability under the effect of nonlinear term $\psi(t,\vv{i_g})$ on line dynamics (\ref{eq:inverter_dynamics_ss_dq}) if (a) the inverter frequency is bounded as $(\Dot{\theta} - \omega_0) \in [-\Delta\omega_{max},\Delta\omega_{max}]$ (b) $G_L S$ is stable and (c) the following bound on the 2-SISO sensitivity $\widetilde{S}$ holds {\small\begin{align}
        \|\widetilde{S}(j\omega)\|_2 &< 
        \frac{\|\Gamma\mathcal{X}_c\|_2^{\text{-}1}}
        {\Delta\omega_{max}}
        \|G_L(j\omega)\|_2^{\text{-}1},
        \quad \forall \omega.
        \label{eq:Nonlinear_Condition}
    \end{align}}
\end{propositionE}
\begin{proofE}
Assume that the line dynamics $G_L$ and the feedback controller $K=\widetilde{K}K_L$, as depicted in Fig. \ref{fig: Simple_Feedback}, are represented by the following state-space realization:
{\small\begin{align}
    \frac{d}{dt}
    x_g
    &=
    A_g
    x_g
    - \psi (t,y_g) +
    y_k,
    &
    y_g &= x_g
    \label{eq:line_model_nonlinear}
    \\
    \frac{d}{dt}x_k &= A_k x_k - B_k y_g, 
    &
    y_k &= C_k x_k + D_k y_g,
    \label{eq:controller_state_space}
\end{align}}%
where $\{A_k,B_k,C_k, D_k\}$ in (\ref{eq:controller_state_space}) represent the minimal realization of controller $K$. Moreover, $A_g$ and $\psi(\cdot,\cdot)$ in (\ref{eq:line_model_nonlinear}) denote the state matrix and the nonlinear dynamic below
{\small\begin{align}
    A_g &= 
    \begin{bmatrix}
        -\frac{R}{L} & \omega_0 \\
        -\omega_0 & -\frac{R}{L}
    \end{bmatrix},
    &
    \psi (t,y_g) &=
    \Delta \omega(t)
    \begin{bmatrix}
        0 & -1 \\
        1 & 0 
    \end{bmatrix}
    \begin{bmatrix}
        y_g^d \\ y_g^q
    \end{bmatrix}.
    \label{eq:nonlinear_element}
\end{align}}%

Our primary objective is to ensure the stability of the closed-loop in Fig. \ref{fig: Simple_Feedback}, represented by the line and controller dynamics in (\ref{eq:line_model_nonlinear}) and (\ref{eq:controller_state_space}), in the presence of nonlinear dynamics $\psi(\cdot,\cdot)$. To establish a stability condition, we first reformulate the dynamics in (\ref{eq:line_model_nonlinear}) and (\ref{eq:controller_state_space}) as the following linear time-invariant (LTI) system
{\small\begin{align}
\begin{split}
    \frac{d}{dt}
    \begin{bmatrix}
        x_g \\
        x_k
    \end{bmatrix} &=
    \begin{bmatrix}
        A_g + D_k & C_k \\
        -B_k & A_k
    \end{bmatrix}
    \begin{bmatrix}
        x_g \\
        x_k
    \end{bmatrix}
    +
    \begin{bmatrix}
        I_2 \\
        \mathbf{0}
    \end{bmatrix}
    u_{\psi},
    \\
    y &=
    \begin{bmatrix}
        I_2 & \mathbf{0} 
    \end{bmatrix}
    \begin{bmatrix}
        x_g \\
        x_k
    \end{bmatrix},
    \label{eq:GLS_Nonlinear}
\end{split}
\end{align}}%
with the nonlinear feedback law  
{\small\begin{align}
\begin{split}
    u_{\psi} &= -\psi(t,y).
    \label{eq:Nlinear_Feedback}
\end{split}
\end{align}}%
By representing the nonlinear line dynamics as the feedback nonlinearity in equation (\ref{eq:Nlinear_Feedback}), we can establish a sufficient stability condition using a sector-boundedness argument.

The nonlinear element $\psi(t,y)$ belongs to the sector $[K_{min}, K_{max}]$ if
{\small\begin{align}
    [\psi(t,y) - K_{min}y]^{\top}
    [\psi(t,y) - K_{max}y] &\leq 0,
    &
    &\forall t.
    \label{eq:sector_definition}
\end{align}}%
Assuming the maximum allowable deviation in the inverter's frequency from the nominal value is bounded as follow
{\small\begin{align}
    \omega_0 - \Delta\omega_{max} &<
    \Dot{\theta}
    < \omega_0 + \Delta\omega_{max},
    \label{eq:freq_deviation}
\end{align}}%
then $K_{max} = I_2 \Delta \omega_{max}$ and $K_{min} = -I_2 \Delta \omega_{max}$ provide one such sector bound for $\psi(\cdot,\cdot)$ in (\ref{eq:nonlinear_element}). Now we are ready to present the stability condition.

Take $G(s)$ as the transfer matrix realization of the state-space model in (\ref{eq:GLS_Nonlinear}), then the sufficient conditions for the stability of the system in (\ref{eq:GLS_Nonlinear}) with the nonlinear feedback law in (\ref{eq:Nlinear_Feedback}) are as follows: (a) $G(s)(I_2 - \Delta\omega_{max}G(s))^{\text{-}1}$ is stable, (b) $Z(s)$ below is strictly positive real (SPR) \cite{khalil2002nonlinear}
{\small\begin{align}
    Z(s) =
    \left(I_2 + \Delta\omega_{max}G(s)\right)
    \left(I_2 - \Delta\omega_{max}G(s)\right)^{\text{-}1}.
    \label{eq:Zs_SPR}
\end{align}}%
We can simplify the stability of $G(s)(I_2 - \Delta\omega_{max}G(s))^{\text{-}1}$ and the SPR condition for $Z(s)$ into the following sufficient conditions, which will be more useful for control design
{\small\begin{align}
    \Delta\omega_{max}
    \|G(j\omega)\|_{\infty} < 1,
    \label{eq:simplified_SPR}
\end{align}}%
and $G(s)$ is stable. We demonstrate the sufficiency of the above condition by examining the three SPR conditions for $Z(s)$, assuming that (\ref{eq:simplified_SPR}) holds \cite{khalil2002nonlinear}.\\
(a) The plot of $\det[I_2 - \Delta\omega_{max}G(j\omega)]$ neither goes through nor circles the origin under the condition in (\ref{eq:simplified_SPR}). Hence, by the multivariate Nyquist criterion, $G(s)(I_2 - \Delta\omega_{max}G(s))^{\text{-}1}$ and subsequently $Z(s)$ is Hurwitz.\\
(b) $Z(j\omega) + Z^\top(-j\omega)$ is positive definite for all $\omega$ if (\ref{eq:simplified_SPR}) holds. Consider the following factorization 
{\small\begin{align}
    \begin{split}
        Z(j\omega) + Z^\top(-j\omega) &= 2
        (I_2 - \Delta\omega_{max}G^\top(-j\omega))^{\text{-}1}
        \\
        &\times
        (I_2 - \Delta\omega_{max}^2G^{\top}(-j\omega)G(j\omega))
        \\
        &\times
        \left(I_2 - \Delta\omega_{max}G(j\omega)\right)^{\text{-}1}
    \end{split}
\end{align}}%
Based on factorization above $Z(j\omega) + Z^\top(-j\omega)>0$ for all $\omega$ if and only if
{\small\begin{align}
    \sigma_{\min}(I_2 - \Delta\omega_{max}^2G^{\top}(-j\omega)G(j\omega)) &> 0, 
    & \forall \omega\in \mathbb{R},
\end{align}}%
where $\sigma_{\min}$ is the smallest singular value. Note that this is indeed the case since $1 - (\Delta\omega_{max}\|G(j\omega)\|_\infty)^2 > 0$ fulfills following lower-bound
{\small\begin{align}
    \sigma_{\min}(I_2 - \Delta\omega_{max}^2G^{\top}(-j\omega)G(j\omega)) > 1 - (\Delta\omega_{max}\|G(j\omega)\|_\infty)^2.
    \nonumber
\end{align}}%
(c) For strictly proper $G(s)$ we have 
{\small\begin{align}
    Z(\infty) + Z^\top(\infty) = 2I_2 > 0.
\end{align}}
To establish practical design criteria for the feedback controller, we need to explicitly express $G(s)$ in (\ref{eq:simplified_SPR}) in terms of the relevant closed-loop transfer functions of Fig. \ref{fig: Simple_Feedback}. This can be done by noting that $G_L S$ has the same state-space realization as (\ref{eq:GLS_Nonlinear}), or simply $G(s) = G_L S$. Consequently, we can convert (\ref{eq:simplified_SPR}) into a condition on the magnitude of the closed-loop sensitivity as follows:
{\small\begin{align}
    \|S(j\omega)\|_2 &< 
    \frac{1}
    {\Delta\omega_{max}}
    \|G_L(j\omega)\|_2^{\text{-}1},
    \quad \forall \omega.
\end{align}}%
As the final step, we replace $S$ with its factorization in terms of $\widetilde{S}$ from Proposition \ref{prop:SISO_Factorization} to get the desired result in (\ref{eq:Nonlinear_Condition}).
\end{proofE}
\section{Specifying Objectives and Operating Modes Based on the Closed-Loop Sensitivity}
\subsection{Sensitivity and Inverter Closed-Loop Objectives}
\label{sec: Extending_the_Steady_State_Analysis}
\label{sec: Control_Objective}
The primary steady-state control objectives of inverters are to keep the output voltage and inverter frequency close to their nominal values and to accurately track the current setpoint. Under ideal conditions, these objectives translate to the following steady-state requirements:
{\small\begin{align}
    \lim_{t\to \infty} \Delta\vv{v_c}(t) &= \bf{0},
    & 
    \lim_{t\to \infty} \Delta\omega(t) &= 0,
    &
    \lim_{t\to \infty} \vv{e}'(t) &= \bf{0}.
    \nonumber
\end{align}}%
In the context of Fig. \ref{fig: Simple_Feedback} and (\ref{eq:control_disturbance}), $\Delta \vv{v_c}$ and $\Delta \omega$ are components of the closed-loop control effort $\vv{u}$, while $\vv{e}'$ represents the modified current feedback error. This observation, together with (\ref{eq:SISO_Effort_Simplified}), enables us to formulate the inverter control objectives in terms of closed-loop {\em sensitivity} $\widetilde{S}$ and {\em complementary sensitivity} $\widetilde{T}$ as follows
{\small\begin{align}
    \|\vv{\hat{u}}\|_2 \leq
    \|\widetilde{T}\|_2
    \|G_M^{\text{-}1}\vv{\hat{i}_0} + \vv{\hat{d}}\|_2,
    \;
    \|\vv{\hat{e}'}\|_2 \leq
    \|\widetilde{S}\|_2
    \|\vv{\hat{i}_0'} + \widetilde{G}_M\vv{\hat{d}}\|_2.
    \label{eq:u_bottleneck}
\end{align}}%
It is desirable to minimize  $\|\widetilde{T}\|_2$ and $\|\widetilde{S}\|_2$ to reduce effect of the exogenous inputs $\{\vv{i_0},\vv{d}\}$ on $\{\vv{u},\vv{e}\}$, thereby improving voltage, frequency, and current regulation. However, the algebraic constraint $\widetilde{S} + \widetilde{T} = I_2$, evident from (\ref{eq:SISO_Complementary_Sensitivity}), indicates a fundamental trade-off. This trade-off exists between achieving perfect voltage and frequency regulation (VSI operation) by keeping $\|\widetilde{S}\|_2$ small and perfect current tracking (GFL operation) by reducing $\|\widetilde{T}\|_2$. The GFM operation is based on adjusting the sensitivity $\widetilde{S}$ to achieve the desired balance between the VSI and GFL operations; as illustrated in Fig. \ref{fig:GFL_to_GFL}.\\
$\bullet$ \textbf{Voltage vs current harmonic trade-off:} An ideal proportional-resonant (PR) compensator with a resonant frequency of $\omega_h$ in $\widetilde{K}$, leads to $\widetilde{S}(j\omega_h) = 0$ and $\widetilde{T}(j\omega_h) = I_2$. This shifts all harmonics from line current to capacitor voltage and inverter frequency according to (\ref{eq:u_bottleneck}). On the other hand, the notch filter in $\widetilde{K}$, results in $\widetilde{S}(j\omega_h) = I_2$ and $\widetilde{T}(j\omega_h) = 0$, shifting the harmonics from voltage and frequency to current.\\
$\bullet$ \textbf{Line-to-ground fault ride-through:} Line-to-ground fault results in a considerable second harmonic component in grid disturbance $\vv{d}$ \cite{askarian2024enhanced}. Overlooked, the disturbance will spread to the line current, causing a fault trip. A second harmonic PR compensator ensures $\widetilde{S}(j2\pi 120)=0$ and $\|\vv{e}(j2\pi 120)\|_2 = 0$ in (\ref{eq:u_bottleneck}), providing perfect control of the line current at 120Hz.
\subsection{Sensitivity and Four Fundamental Operating Modes}
The inverter's operation mode is determined by how the sensitivity $\widetilde{S}$ shapes and mitigates the impact of grid disturbance $\vv{d}$ on the current feedback error $\vv{e}'$ during steady-state operation. To illustrate this, consider the transfer function between the grid disturbance $\vv{d}$ and $\vv{e}'$ from (\ref{eq:SISO_Effort_Simplified}) and expand $\vv{d}$ according to the nominal disturbance model in (\ref{eq:disturbance_type}) to get
{\small\begin{align}
    \begin{bmatrix}
        \hat{e}'^d \\ \hat{e}'^q
    \end{bmatrix} =
    \widetilde{G}_M
    \widetilde{S} \vv{\hat{d}}
    \approx
    \widetilde{G}_M
    \begin{bmatrix}
        \widetilde{S}^d & 0 
        \\
        0 & \widetilde{S}^q
    \end{bmatrix}
    \begin{bmatrix}
            1/s & 0\\
            0 & v_0/s^2
        \end{bmatrix}
        \begin{bmatrix}
            \Delta v_g \\
            \Delta\omega_g
        \end{bmatrix},
        \label{eq:Droop_Base_equation}
\end{align}}%
Based on (\ref{eq:Droop_Base_equation}), there are only four meaningful scenarios for steady-state operation. These scenarios depend on whether $\widetilde{S}^d$ and $\widetilde{S}^q$ can completely reject grid disturbances $\{\Delta v_g,\Delta \omega_g\}$ in (\ref{eq:Droop_Base_equation}) or result in a finite gain between $\{\Delta v_g,\Delta \omega_g\}$ and $\{e'^d,e'^q\}$ during steady-state operation. In the following proposition, we demonstrate that these scenarios correspond to GFL, GFM, ESS, and STATCOM operation, and can be characterized by the zeros of $\widetilde{S}^d$ and $\widetilde{S}^q$.
\begin{propositionE}
\label{sec:GFL_GFM}
Inverter operating modes are determined by the number of zeros at the origin in $\widetilde{S}^d$ and $\widetilde{S}^q$ as follow:\\
(a) \textbf{GFL:} If $\widetilde{S}^d$ has one zero and $\widetilde{S}^q$ includes two zeros at the origin, we achieve complete rejection of grid disturbances, and
{\small\begin{align}
    \lim_{t \to \infty} e'^d &= 0,
    &
    \lim_{t \to \infty} e'^q &= 0.
\end{align}}%
(b) \textbf{STATCOM:} If $\widetilde{S}^d$ has no zero and $\widetilde{S}^q$ includes two zeros at the origin, the inverter operates as STATCOM and
{\small\begin{align}
    \lim_{t \to \infty} e'^d 
    &= \Delta v_g / 
    (\widetilde{G}_M^{\text{-}1}(j0) + K^d(j0)),
    &
    \lim_{t \to \infty} e'^q &= 0.
\end{align}}%
(c) \textbf{ESS:} If $\widetilde{S}^d$ and $\widetilde{S}^q$ each maintain one zero at the origin, the inverter functions as ESS and
{\small\begin{align}
    \lim_{t \to \infty} e'^d &= 0,
    &
    \lim_{t \to \infty} e'^q
    &=
    \lim_{s \to 0} \Delta \omega_g/(s K^q(s)).
\end{align}}%
(d) \textbf{GFM:} If $\widetilde{S}^d$ has no zero at the origin and $\widetilde{S}^q$ has one zero at the origin, the closed-loop in Fig. \ref{fig: Simple_Feedback} leads to an inherent and well-posed drooping characteristic in steady state given below
{\small\begin{align}
    \begin{bmatrix}
        \Delta v_g \\ \Delta \omega_g
    \end{bmatrix} =
    \begin{bmatrix}
        \widetilde{G}_M^{\text{-}1}(j0) + K^d(j0) & 0
        \\
        0 & \lim_{s \to 0} sK^q(s)
    \end{bmatrix}
    \begin{bmatrix}
        e'^d \\ e'^q
    \end{bmatrix}.
    \label{eq:implicit_droop}
\end{align}}%
\end{propositionE}
\begin{proofE}
    (a), (b), (c) and (d) are the direct result of application of the final value theorem to (\ref{eq:Droop_Base_equation}). $\blacksquare$
\end{proofE}

\begin{figure}[t]
    \captionsetup[subfigure]{labelformat=empty}
    \centering
    \subfloat[]{\includegraphics[width = 0.51\columnwidth]{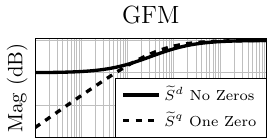}}
    \hfill
    \subfloat[]{\includegraphics[width = 0.45\columnwidth]{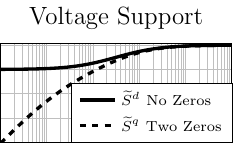}}
    \\
    \vspace{-7mm}
    \subfloat[]{\includegraphics[width = 0.515\columnwidth]{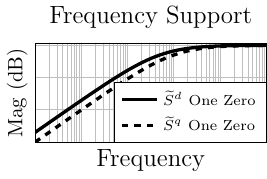}}
    \hfill
    \subfloat[]{\includegraphics[width = 0.445\columnwidth]{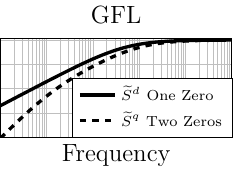}}
    \vspace{-5mm}
    \caption{Inverter operating mode based on the shape and number of zeros in $\widetilde{S}$ at low-frequency. (a) GFM, (b) voltage support (STATCOM), (c) frequency support (ESS), and (d) GFL.}
    \label{fig:Four_Inverter}
\end{figure}

\begin{remark}
    Fig. \ref{fig:Four_Inverter} shows the operating modes based on the shape of $\widetilde{S}$ at low frequencies. For control design, we can leverage the definition of $\widetilde{S}$ in (\ref{eq:SISO_Sensitivity}) to convert the conditions on the number of zeros, as outlined in Proposition \ref{sec:GFL_GFM}, into an equivalent condition regarding the number of integrators in $K^d$ and $K^q$. This relationship is summarized in Table \ref{tab:Mode_Integral}.
\end{remark}
\begin{remark}
    The more accurate term for STATCOM and ESS operation in the Proposition \ref{sec:GFL_GFM} is the voltage support and the frequency support, since in steady state $e'^d$ and $e'^q$ are linear combinations of $e^d$ and $e^q$ specified by $K_L(j0)$. For example, with $K_L$ in (\ref{eq:KL_Diagonal}) we have 
    {\small\begin{align}
        \vv{e}' = K_L \vv{e} 
        \xRightarrow[]
        {\text{\normalsize (\ref{eq:KL_Diagonal})}}
        \begin{bmatrix}
            e'^d \\ e'^q
        \end{bmatrix} =
        \begin{bmatrix}
            \cos{\phi_z} & -\sin{\phi_z}
            \\
            \sin{\phi_z} & \hphantom{-}\cos{\phi_z}
        \end{bmatrix}
        \begin{bmatrix}
            e^d \\ e^q
        \end{bmatrix}.
    \end{align}}%
    In this case, the nature of power transactions for voltage and frequency regulation is a combination of active and reactive power, depending on the characteristics of the line. For inductive lines ($\phi_z \approx 90^\circ$), we observe conventional STATCOM and ESS operation, where voltage regulation is achieved through reactive current transactions and frequency regulation through active current transactions.
\end{remark}
\begin{table}
\fbox{
\begin{minipage}{0.54\columnwidth}
    \captionsetup{font={default}}
    \centering
    \resizebox{\linewidth}{!}{%
    \begin{tabular}{c|c|c}
         \diagbox[height = 0.75cm]{$K^d$}{$K^q$}& $\mathbf{1} \times (1/s)$ 
         & $\mathbf{2} \times (1/s)$  \\
         \hline
         $\mathbf{0} \times (1/s)$ & GFM & ESS \\
         \hline
         $\mathbf{1} \times (1/s)$ & STATCOM & GFL \\
         \hline
    \end{tabular}}
    \captionof{table}{\normalsize Inverter modes and the number of integrators ($1/s$) in $\{K^d,K^q\}$.}
    \label{tab:Mode_Integral}
\end{minipage}}
\hfill
\fbox{
\begin{minipage}{0.37\columnwidth}
    \captionsetup{font={default}}
    \centering
    \includegraphics[width = 1\columnwidth]{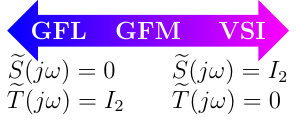}
    \captionof{figure}{\normalsize Spectrum between GFL and VSI parameterized by $\widetilde{S}$.}
    \label{fig:GFL_to_GFL}
\end{minipage}}
\vspace{-1.5em}
\end{table}%
\subsection{Power-Sharing Among Parallel GFM Inverters} 
Power sharing between parallel inverters is essential for the sustained operation of GFM inverters with varying power ratings and to enforce the DER usage priority \cite{baranwal2018distributed}. Power sharing can be defined as the change in the active and reactive output power or current from the nominal setpoint, proportional to changes in the PCC voltage $\Delta v_g$ and frequency $\Delta \omega_g$. This proportion for each GFM inverter determines the sharing ratio. Based on this definition and the droop law in (\ref{eq:implicit_droop}) the sharing ratio between two arbitrary GFM inverters $(i)$ and $(j)$ is
{\small\begin{align}
    \frac{(\widetilde{G}_M^{\text{-}1}(j0) + 
    K^{d}(j0))^{(i)}}
    {(\widetilde{G}_M^{\text{-}1}(j0) + 
    K^{d}(j0))^{(j)}}
    &=
    \frac{e'^{d(j)}}{e'^{d(i)}},
    &
    \lim_{s \to 0}
    \frac{K^{q}(s)^{(i)}}
    {K^{q}(s)^{(j)}}
    &=
    \frac{e'^{q(j)}}{e'^{q(i)}}.
    \label{eq:sharing_decoupled}
\end{align}}%
\begin{remark}
    $d$-axis sharing in (\ref{eq:sharing_decoupled}) depends on the controller $K^d$ and the line impedance $\widetilde{G}_M^{\text{-}1}$. However, we can reduce the effect of line impedance on power sharing and droop behavior in (\ref{eq:implicit_droop}) by increasing the DC gain of the controller $\widetilde{G}_M^{\text{-}1}(j0) \ll K^d(j0)$. This subsumes the concept of virtual impedance. 
\end{remark}
\begin{remark}
\label{rem:exact_sharing}
    The $q$-axis sharing in (\ref{eq:sharing_decoupled}), unlike the $d$-axis, depends solely on $K^q$. This allows for exact sharing of the $e'^q$ component between inverters with different line impedances. However, based on the specific choice of $K_L$, $e'^q$ in (\ref{eq:sharing_decoupled}) represents a linear combination of $e^d$ and $e^q$. For example, $K_L(j0)$ in (\ref{eq:KL_Diagonal}) results in $e'^q = [\sin{\phi_z}\; \cos{\phi_z}] \vv{e}$. This particular choice of $K_L$ leads to exact active power sharing only for inductive lines ($\phi_z \approx 90^0$). However, we can achieve exact active current sharing by adopting $K_L^*$ in (\ref{eq:KL_Trian}), which maps $e'^q$ to $e^d$ in steady state, regardless of line impedance.
    {\begin{align}
        K_L^{*}(s) =
        &\frac{\omega_m}{s + \omega_m}
        \begin{bmatrix}
            \frac{L}{Z}s + \cos{\phi_z} & 
            -\sin{\phi_z}
            \\
            \frac{\lambda}{\lambda^2 + \omega_0^2} s + 1 &
            \frac{\omega_0}{\lambda^2 + \omega_0^2} s
        \end{bmatrix},
        \label{eq:KL_Trian}
    \end{align}}%
    The proposed $K_L^*$ satisfies stability criteria outlined in parts (a) and (b) of the Proposition \ref{prop: rotation_stability}, resulting in the following modified plant $G_M^*$ and its diagonal counterpart $\widetilde{G}_M^*$ below
    {\small\begin{align}
        G_M^{*} &=
        \frac{\omega_m/Z}{s + \omega_m}
        \begin{bmatrix}
            1 & 0 \\
            \cos{\phi_z} & \sin{\phi_z}
        \end{bmatrix},
        &
        \widetilde{G}_M^{*} &=
        \frac{\omega_m/Z}{s + \omega_m}
        \begin{bmatrix}
            1 & 0 \\
            0 & \sin{\phi_z}
        \end{bmatrix}.
        \label{eq:GM_Modified_Two}
    \end{align}}%   
    However, $G_M^{*}(j0)$ in (\ref{eq:GM_Modified_Two}) has a suboptimal condition number of $\kappa\left(G_M^*(j0)\right) = (1 + \cos{\left(\phi_z\right)})/\sin{\left(\phi_z\right)}$ and for $\phi_z \leq 11^\circ$, (dominantly resistive line), the condition number exceeds $10$. This suggests that $K_L^*$ trades the robustness margin to achieve exact active power sharing.
\end{remark}%
\begin{remark}
    Based on (\ref{eq:implicit_droop}) the steady-state gain of the $d$ and $q$ axes directly influences the drooping characteristics of the inverter in GFM mode. For applications requiring small drooping coefficients, the controllers' low-frequency gain will be limited. Consequently, recalling the definition of sensitivity $\widetilde{S}$ from (\ref{eq:SISO_Sensitivity}), we cannot arbitrarily reduce the sensitivity to satisfy the stability condition in part (c) of Proposition \ref{prop:MIMO_Stability}. This observation justifies the need for the stability conditions provided in part (b) of Proposition \ref{prop:MIMO_Stability}.
\end{remark}
\begin{figure}[t]
    \centering
    \subfloat[]{\includegraphics[width = 0.85\linewidth]{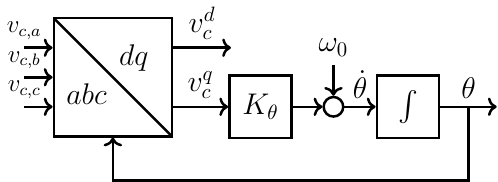}
    \label{fig:PLL_Conventional}}
    \vspace{-1mm}
    \subfloat[]{\includegraphics[width = 0.85\linewidth]{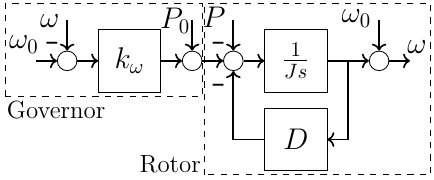}
    \label{fig:Governor_Rotor}}
    \caption{(a) Conventional 3-phase PLL system for GFL inverters that include $abc$ to $dq$ transform, PLL loop filter $K_\theta$ and voltage controlled oscillator (VCO) (integrator). (b) Modified frequency control for GFM inverters to mimic rotor dynamics and governor action.}
\end{figure}%
\subsection{Synchronization and Frequency Transient Response}
Existing GFL and STATCOM use the PLL mechanism such as one show in Fig. \ref{fig:PLL_Conventional} to synchronize their rotating frame with an established grid voltage. However, GFM and ESS synchronize through power transactions based on a frequency droop scheme such as a virtual inertia frame that is shown in Fig. \ref{fig:Governor_Rotor}. Therefore, changing between the GFL/STATCOM and GFM/ESS modes involves changing the synchronization method. In this section, we exploit the parallel structure of the $q$ axis controller, that is $K^q = K_1^q + K_2^q$, in Fig. \ref{fig: Simple_Feedback} to develop a mode-independent synchronization scheme and generalize the concept of inertia and rate-of-change of frequency (RoCoF) as a closed-loop transient response. 

In the context of the closed-loop signals in Fig. \ref{fig: Simple_Feedback} the two primary synchronization conditions are
{\small\begin{align}
    \text{\normalsize (a)}
    \lim_{t\to \infty} \Delta v_c^q = 0,
    \;\text{\normalsize (b)}
    \lim_{t\to \infty} v_0(\Dot{\theta}_g - \Dot{\theta}) 
    \approx
    \lim_{t\to \infty} \Dot{d}^q - \Dot{u}_\theta = 0,
    \label{eq:Synch_Condition}
\end{align}}%
where the approximation in (b) is based on definition of $d^q$ and $u_\theta$ in Proposition \ref{prop:control_disturbance}. Condition (a) aligns the capacitor voltage phasor $\vv{v}_c$ with the $dq$ frame while (b) ensures that the rotating frame achieves frequency lock with the grid frequency $\Dot{\theta}_g$. Note that condition (a) in (\ref{eq:Synch_Condition}), that is $v_c^q=0$, is necessary to decouple the following mapping between the active and reactive power and $dq$ current
{\small\begin{equation}
    \begin{split}
        \begin{bmatrix}
        i_0^d \\ i_0^q
        \end{bmatrix} = 
        \frac{\Phi_{\{1, 3\}}}{\|\vv{v_c}(t)\|^2}
        \begin{bmatrix}
        v_c^d & \hphantom{-}v_c^q \\
        v_c^q & -v_c^d
        \end{bmatrix}
        \begin{bmatrix}
        P_0 \\ Q_0
        \end{bmatrix}.
    \end{split}
    \label{eq:power_current_mapping}
\end{equation}}%
Here, $\Phi_1 = 1$ for single and $\Phi_3 = 2/3$ for three-phase system.

Both conditions in (\ref{eq:Synch_Condition}) are expressed in terms of the closed-loop variables $\Delta v_c^q$ and $u_\theta$, along with the grid disturbance $d^q$ (see Fig. \ref{fig: Simple_Feedback}). This formulation enables us to translate the synchronization conditions in (\ref{eq:Synch_Condition}) into constraints for the feedback control design. To proceed, we first derive the transfer function between the exogenous inputs $i_0'^q$ and $d^q$, and $\Delta v_c^q$ and $u_\theta$ based on Fig. \ref{fig: Simple_Feedback} and (\ref{eq:SISO_Effort_Simplified}) as follows:
{\small\begin{align}
\begin{split}
    \Delta \hat{v}_c^q &= K_1^q \hat{e}'^{q} = 
    K_1^q \widetilde{S}^q(\hat{i}_0'^{q} + \widetilde{G}_M^q \hat{d}^q),
    \\
    \hat{u}_\theta &= K_2^q \hat{e}'^{q} = 
    K_2^q \widetilde{S}^q(\hat{i}_0'^{q} + \widetilde{G}_M^q \hat{d}^q).
\end{split}
\label{eq:Internal_control_effort}
\end{align}}%
Defining $T_v$ and $T_\theta$ transfer functions as follow
{\small\begin{align}
    T_v &=
    \frac{G_M^qK_1^q}
    {1 + G_M^q(K_1^q + K_2^q)},
    &
    T_\theta &=
    \frac{G_M^q K_2^q}
    {1 + G_M^q\left(K_1^q + K_2^q\right)},
    \label{eq:Freq_TF}
\end{align}}%
and $\hat{w} = (\widetilde{G}_M^{q})^{\text{-}1}\hat{i}_0'^q$ we convert (\ref{eq:Internal_control_effort}) into following form
{\small\begin{align}
    \Delta \hat{v}_c^q &= T_v
    ( \hat{w} + \hat{d}^q ),
    &
    \hat{u}_\theta &= T_\theta
    ( \hat{w} + \hat{d}^q ),
    \label{eq:final_value_synchronization}
\end{align}}%
By applying the final value theorem to (\ref{eq:Synch_Condition}) and substituting $\Delta v_c^q$ and $u_\theta$ from (\ref{eq:final_value_synchronization}), we can derive the synchronization conditions in parts (a) and (b) of (\ref{eq:Synch_Condition}) in the following practical form:
{\small\begin{align}
    \lim_{s \to 0} s T_v (
    \hat{w} +
    \hat{d}^q)&= 0,
    &
    \lim_{s \to 0} s^2\left( (1 - T_\theta)\hat{d}^q 
    - T_\theta \hat{w} \right) &= 0.
    \label{eq:final_value_complete}
\end{align}}%
The equation above allows us to formulate the synchronization in terms of the conditions on the controllers $K_1^q$ and $K_2^q$.
\begin{propositionE}
\label{Prop: Synchronization}
We achieve synchronization, regardless of the operating mode, if $(K_1^q/K_2^q)$ includes at least two zeros at the origin and $K_2^q$ has at least one pole at the origin.
\end{propositionE}
    \begin{proofE}
        Adopting the nominal ramp model in (\ref{eq:disturbance_type}) for $d^q$ and considering $T_v = T_\theta K_1^q/K_2^q$ we rewrite (\ref{eq:final_value_complete}) as follow
        {\small\begin{align}
        \lim_{s \to 0} s 
        \frac{K_1^q}{K_2^q} T_\theta (
        \hat{w} +
        v_0\frac{\Delta \omega_g}{s^2}) &= 0,
        \,
        \lim_{s \to 0} v_0 (1 - T_\theta)\Delta\omega_g 
        - s^2 T_\theta \hat{w} = 0.
        \nonumber
    \end{align}}%
    Based on the conditions in the proposition $\lim_{s \to 0} T_\theta = 1$ and $\lim_{s\to 0} K_1^q/(s^2 K_2^q) = \text{\em const}$, completing the proof. 
    $\blacksquare$
\end{proofE}

\subsubsection{Transient response of Inverter Frequency}
\label{sec:high_freq_correction}
$T_\theta$ in (\ref{eq:Freq_TF}) represents the closed-loop transfer function between the inverter's frequency deviation $\Delta \omega$, and the grid frequency deviation $\Delta \omega_g$. Therefore, the characteristics of $T_\theta$, such as the bandwidth, generalize the concepts of inertia and RoCoF, while the peak of $T_\theta$, denoted as $M_T = \|T_\theta\|_\infty$, is directly proportional to the overshoot (frequency nadir) and the transient oscillation of $\Delta \omega$. Therefore, it is desired to shape $T_\theta$ as a unity gain low-pass filter with $M_T < 3$dB \cite{skogestad2005multivariable}, and low bandwidth. However, there are two sources of difficulties. First, the transient frequency response, the current and the voltage regulation objectives on the $q$ axis are interdependent and conflicting. Second, shaping $T_\theta$ and $T_v$ in (\ref{eq:Freq_TF}) to achieve the desired closed-loop objectives requires the concurrent design of both $K_1^q$ and $K_2^q$.

\begin{figure}[t]
    \centering
    \includegraphics[width= 0.95\linewidth]{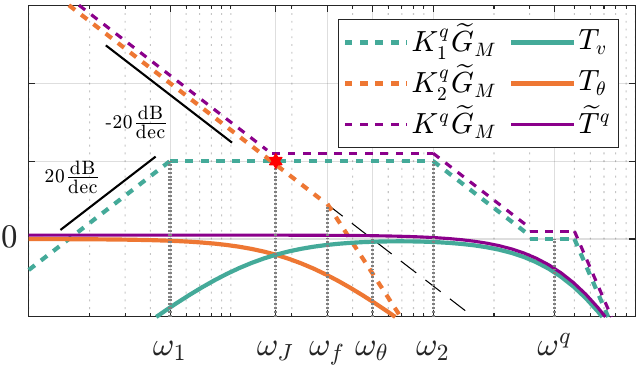}
    \caption{Outline of desired $q$ axis closed-loops $T_v$, $T_\theta$, $\widetilde{T}^q$, and corresponding open-loops $K_1^q\widetilde{G}_M$, $K_2^q\widetilde{G}_M$, and $K^q\widetilde{G}_M$. }
    \label{fig:Desired_Controller}
    \vspace{-3mm}
\end{figure}

We establish the design criteria for $K_1^q$ and $K_2^q$ considering the fundamental algebraic constraints outlined below
{\small\begin{align}
    T_\theta + T_v &=
    \widetilde{T}^q,
    &
    \widetilde{S}^q + \widetilde{T}^q
    &= \widetilde{S}^q + (T_\theta + T_v^q)
    =1.
    \label{eq:S_T_q}
\end{align}}%
$\widetilde{T}^q$ is the closed-loop between the current setpoint $i_0'^q$ and $i_g'^q$. We aim to keep $\widetilde{T}^q$ close to one and $\widetilde{S}^q$ small over a broad frequency range $[0, \omega^q]$ to achieve effective disturbance rejection (\ref{eq:SISO_Effort_Simplified}), negligible tracking error (\ref{eq:u_bottleneck}), dynamic decoupling (\ref{eq:magnitude_condition}), and MIMO stability (part (c) of Proposition \ref{prop:MIMO_Stability}). Furthermore, $T_\theta$ in (\ref{eq:Freq_TF}) should be designed as a low-pass filter with a bandwidth $\omega_J \ll \omega^q$ to achieve an inertial response, low RoCoF, and to attenuate the effect of high-frequency grid disturbances on $\Delta \omega$. Based on (\ref{eq:S_T_q}), this automatically constrains $T_v$ to a unity gain band-pass filter between $\omega_J$ and $\omega_q$ as shown in Fig. \ref{fig:Desired_Controller}. In the following, we sketch the outline of the open-loops $K_1^q\widetilde{G}_M$ and $K_2^q\widetilde{G}_M$ that result in the desired $T_\theta$, $T_v$ and $\widetilde{T}^q$. Subsequently, in Section \ref{subsec:Feedback_Control_Design}, we provide an example control design for $K_1^q$ and $K_2^q$.%

\textbf{$K_1^q\widetilde{G}_M$:} should be designed as a bandpass filter with a gain cross-over frequency near $\omega_q$ and a phase margin of $60^\circ$. The pass band should satisfy $\omega_J\in[\omega_1,\omega_2]$, as shown in Fig. \ref{fig:Desired_Controller}, where $\omega_J$ is the bandwidth of $T_\theta$ and adjusts the desired inertial response. The ratio $\omega_J/\omega_1$ is directly proportional to $M_T$ and damps and reduces the frequency nadir and oscillation. Finally, the filter should achieve a slope of $-20$dB/dec at low frequency to satisfy the synchronization condition in Proposition \ref{Prop: Synchronization}.

\textbf{$K_2^q\widetilde{G}_M$:} should be shaped close to an integrator (e.g. $\omega_\theta/s$) that intersects $K_1^q\widetilde{G}_M$ at $\omega_J$ in the pass band region, as shown in Fig. \ref{fig:Desired_Controller}. To attenuate the effect of high-frequency disturbances on inverter frequency, we augment the open-loop with a low-pass filter with cutoff frequency $\omega_f$ above $\omega_J$.
\section{Control Design and Implementation}
\label{sec: Control_Design}
In this section, we provide an example of feedback control design for $\widetilde{K}$ that meets the stability and performance criteria proposed in this paper. Subsequently, we implement the proposed control design using inverter closed-loop dynamics.
\subsection{Feedback Control Design}
\label{subsec:Feedback_Control_Design}
For the control design, we assume the nominal line impedance of $Z\angle \phi_z$, and adopt $K_L$ in (\ref{eq:KL_Diagonal}) and the corresponding $\widetilde{G}_M$ in (\ref{eq:GM_Modified}) to satisfy the coupling stability in Proposition \ref{prop: rotation_stability}. Subsequently, for $\widetilde{K}= \text{diag}  (K^d,K_1^q + K_2^q)$ we employ the following set of controllers
{\small\begin{align}
    K^d &=
    \frac{s + \omega_m}{\omega_m / Z}
    \left(
    \frac{s + \alpha_v}{s + 2\sqrt{2}(aZ)\beta_v}
    \right)
    \left(
    \frac{\sqrt{2}\omega_d}{s + \omega_d}
    \right)^3
    \left(
    \frac{s + a\omega_d}{as + \omega_d}
    \right),
    \nonumber
    \\
    K_1^q &=
    \frac{s + \omega_m}{\omega_m / Z}
    \left(
    \frac{\sqrt{\omega_q^2 + \omega_2^2} s}
    {(s + \omega_1) (s + \omega_2)}
    \right)
    \left(
    \frac{\sqrt{2}\omega_q}{s + \omega_q}
    \right)^2
    \left(
    \frac{s + a\omega_q}{as + \omega_q}
    \right),
    \nonumber
    \\
    K_2^q &= 
    \frac{s + \omega_m}{\omega_m / Z}
    \left(
    \frac{\omega_\theta}{s}
    \right)
    \left(
    \frac{s + \alpha_\theta /  \omega_\theta}{s + \beta_\theta Z}
    \right)
    \left(
    \frac{\omega_{f}}{s + \omega_{f}}
    \right).
    \label{eq:Sample_Controller}
\end{align}}%
We justify the proposed control parameters and design by considering their dynamic and steady-state characteristics.
\subsubsection{\textbf{\texorpdfstring{$d$}{TEXT}-axis}}
\label{sec: closed-loop_steady_state_droop}
{\em Droop characteristics:} Based on (\ref{eq:implicit_droop}), the steady-state voltage droop coefficient for $K^d$ in (\ref{eq:Sample_Controller}) is
{\small\begin{align}
    \widetilde{G}_M^{\text{-}1}
    + K^d(j0)
    &= Z + \frac{\alpha_v}{\beta_v},
    & 
    \max \left\{
    \alpha_v,\frac{4(aZ)}{\sqrt{2}}\beta_v
    \right\} 
    &\ll \omega_d.
    \label{eq:Kd_Droop}
\end{align}}%
The voltage droop depends on the ratio $\alpha_v/\beta_v$. Higher values of $\alpha_v$ and $\beta_v$ correspond to faster zero and pole in (\ref{eq:Sample_Controller}) and quicker droop responses. We recommend limiting $\alpha_v$ and $\beta_v$ by the $d$-axis open-loop bandwidth as indicated in (\ref{eq:Kd_Droop}).

{\em Dynamic characteristics:}
Given that inequality in (\ref{eq:Kd_Droop}) holds, $K^d$ in (\ref{eq:Sample_Controller}) leads to an open-loop $K^d G_M^d$ with a gain crossover frequency of close to $\omega_d$ and phase margin
{\small\begin{align}
    \text{\normalsize PM} &\approx
    45 + \arcsin{(1 - a) / (1 + a)},
    &
    0 \leq a & \leq 1.
\end{align}}%
\subsubsection{\textbf{\texorpdfstring{$q$}{TEXT}-axis}}
{\em Droop characteristics:} Based on (\ref{eq:implicit_droop}), the steady-state frequency droop coefficient for $K^q$ in (\ref{eq:Sample_Controller}) is
{\small\begin{align}
    \lim_{s \to 0} s(K_1^q + K_2^q) 
    &= \frac{\alpha_\theta}{\beta_\theta},
    &
    \max \left\{
    \frac{\alpha_\theta}{\omega_\theta}, \beta_\theta Z
    \right\} 
    & \ll \omega_J.
    \label{eq:Kq_Droop}
\end{align}}%
We limit $\alpha_\theta$ and $\beta_\theta$ by $\omega_J$ in (\ref{eq:Kq_Droop}), where $\omega_J$ represents the bandwidth for $T_\theta$ to achieve desired inertia.

{\em Dynamic characteristics} $K_1^q$ and $K_2^q$ in (\ref{eq:Sample_Controller}) satisfy the synchronization condition in Proposition \ref{Prop: Synchronization}. Furthermore, for $\omega_1 < \omega_2 \ll \omega^q$ the shape of $K_1^q\widetilde{G}_M$ traces the desired shape shown in Fig. \ref{fig:Desired_Controller}, with a gain crossover frequency of $\omega^q$ and a phase margin close to PM$\approx\arcsin{(1 - a) / (1 + a)}$.\\
The proposed $K_2^q$ in (\ref{eq:Sample_Controller}), includes an integrator $\omega_\theta / s$, low-pass filter with a cutoff frequency of $\omega_f$ ($\omega_J < \omega_f$) to attenuate the disturbance in the inverter frequency, and a compensator to shape the drooping characteristics. The resulting $K_2^q\widetilde{G}_M$ intersects the pass band region of $K_1^q\widetilde{G}_M$ at $\omega_J\in[\omega_1,\omega_2]$ provided that $\omega_\theta$ satisfies the constraint
{\small\begin{align}
    \log(\omega_J) - \log(\omega_2)
    &\approx
    \log(\omega_\theta) - \log(a\omega_q),
    \Leftrightarrow
    \omega_\theta \approx 
    \frac{\omega_q}{a\omega_2} \omega_J.
\end{align}}%
This allows us to calculate $\omega_\theta$ for the desired inertial bandwidth $\omega_J$. Finally, by adjusting $\omega_1$, we can fine-tune the ratio of $\omega_J/\omega_1$ for the desired damping of the frequency oscillation.

\subsubsection{PR compensator for enhanced stability} 
\label{sec: Nonlinear_PR_Design}
Proposition \ref{prop:nonlinear} imposes an upper-bound on the 2-SISO sensitivity $\widetilde{S}$ to ensure stability in the presence of the nonlinear term $\psi(\cdot,\cdot)$ in line dynamics.
\begin{wrapfigure}{r}{0.5\linewidth}
    \includegraphics[width=1\linewidth]{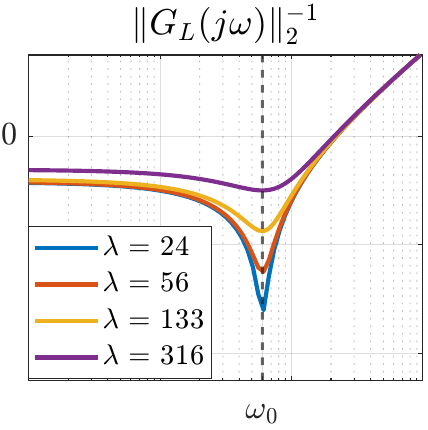}
    \caption{Smaller values of $\lambda$ impose stricter upper-bound on sensitivity around the fundamental frequency $\omega_0$.}
    \label{fig:PR_for_Nlinear}
    \vspace{13mm}
\end{wrapfigure}
This upper bound is explicitly dependent on $\|G_L(j\omega)\|^{\text{-}1}$, where the shape of $G_L(j\omega)$ is uniquely determined by the line characteristics, denoted by $\lambda = R/L$. As shown in Fig. \ref{fig:PR_for_Nlinear}, smaller values of $\lambda$ result in a significant notch in $\|G_L(j\omega)\|^{\text{-}1}$ at the fundamental frequency $\omega_0$. Consequently, this requires $\widetilde{S}$ to be small at $\omega_0$. Based on the definition of $\widetilde{S}$, we can reduce $\widetilde{S}$ at $\omega_0$ by including a proportional-resonant compensator (PR) in series with the $K^d$ and $K_1^q$ controllers in (\ref{eq:Sample_Controller}). An example of a PR compensator is given below
{\small\begin{align}
    K_{\text{PR}} &= 1 +
    k_{\omega_0}\frac{2\zeta\omega_0 s}
    {s^2 + 2\zeta\omega_0 s + \omega_0^2},
    &
    \zeta = \frac{\Delta\omega_{max}}{\omega_0},
\end{align}}%
where $k_{\omega_0}$ sets the PR gain and $\Delta\omega_{max}$ represents the maximum expected frequency deviation from nominal value.

We simplified closed-loop analysis and design neglecting the dependence of $K^d$ and $K_1^q$ on the inverter closed-loop dynamics. In the following section, we discuss how to use inverter dynamics to implement $K^d$ and $K_1^q$.
\subsection{Inverter Closed-Loop as Feedback Compensator}\label{sec: mdesign}
\begin{figure}[t]
    \centering
    \subfloat[]{
    \includegraphics[width = 0.85\columnwidth]{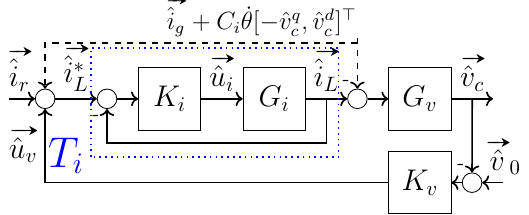}
    \label{fig:Inner_Outer}
    }
    \vspace{-3mm}
    \subfloat[]{
    \includegraphics[width = 0.85\columnwidth]{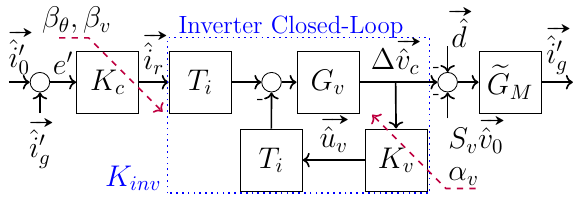}
    \label{fig:Equivalent_Feedback}
    }
    \caption{(a) Nested structure with voltage and current inputs. (b) Inverter closed-loop as part of feedback compensator.}
    \label{fig:Inner_Outer_Equivalent}
    \vspace{-5mm}
\end{figure}

We adopt the cascaded closed-loop structure in Fig. \ref{fig:Inner_Outer} with inner current and outer voltage loops. Here, $G_i = 1/\left(L_i s + R_i\right)$ and $G_v = 1/\left(C_i s\right)$ represent the dynamics of the inductor and capacitor in (\ref{eq:inverter_ss_dynamics}). To form the inner closed-loop, we use feedback linearization to decouple the inductor dynamics in (\ref{eq:inverter_ss_dynamics}) by defining the modulation signal as 
{\small\begin{equation}
    \begin{split}
        \vv{m} = \frac{2}{v_{dc}} 
        (\vv{u_i} + \vv{v_c} + L_i \Dot{\theta}
        [
        -i_L^q, i_L^d
        ]^{\top} ),
    \end{split}
    \label{eq:feedback_linearization}
\end{equation}}%
where $\vv{u_i}$ denotes the output of current controller $K_i$ in Fig. \ref{fig:Inner_Outer}. Subsequently, we use the PI controller $K_i = \omega_c\left(L_i s + R_i\right)/s$ to shape the inner closed-loop $T_i$ in Fig. \ref{fig:Inner_Outer} into a unity gain low-pass filter with cutoff frequency $\omega_c$ below
{\small\begin{equation}
    \begin{split}
        T_i = \frac{G_iK_i}{1 + G_iK_i}=\frac{\omega_c}{s + \omega_c},
        \; \text{and} \;
        S_i = 1 - T_i = \frac{s}{s + \omega_c}.
    \end{split}
    \label{eq: Ti_Si}
\end{equation}}%
The outer voltage loop is formed by setting the reference, $\vv{i^*_L}$, for the inner closed-loop as follow
{\small\begin{equation}
    \begin{split}
        \vv{i_L^*} = \vv{i_{r}}  
        + \vv{u_v} + \vv{i_{g}}
        + C_i \Dot{\theta} [-v_c^q,v_c^d]^{\top},
    \end{split}
\end{equation}}%
where $\vv{u_v}$ is the output of the voltage compensator $K_v$ and $\vv{i_r}$ is the input to the cascaded closed-loop as shown in Fig. \ref{fig:Inner_Outer}. The feedforward term $\vv{\hat{i}_g}+C_i \Dot{\theta} [-v_c^q,v_c^d]^{\top}$ decouples the capacitor dynamics and counteracts the effect of grid current in (\ref{eq:inverter_ss_dynamics}). The proposed cascaded closed-loop in Fig. \ref{fig:Inner_Outer} leads to the following relation between the inputs $\{\vv{v_0},\vv{i_r}, \vv{i_g}\}$ and $\vv{\hat{v}_c}$ 
{\small\begin{equation}
    \begin{split}
        \vv{\hat{v}_c} 
        &=  
        G_v S_v 
        \left(
        T_i \vv{\hat{i}_{r}} +
        T_i K_v \vv{\hat{v}_0}
        - S_i 
        \left(
        \vv{\hat{i}_g} + C_i \Dot{\theta} [-v_c^q,v_c^d ]^{\top}
        \right)
        \right),
        \label{eq:inner_outer_loop}
    \end{split}
\end{equation}}%
where $\{S_i, T_i\}$ are given by (\ref{eq: Ti_Si}), and $S_v$ is the sensitivity transfer function for the voltage closed-loop defined below
{\small\begin{equation}
    \begin{split}
        S_v &=
        \begin{bmatrix}
        S_v^d & 0 \\ 0 & S_v^q
        \end{bmatrix}=
        \begin{bmatrix}
        (1 + G_v T_i K_v^d)^{\text{-}1} & 0\\
        0 & (1 + G_v T_i K_v^q)^{\text{-}1}
        \end{bmatrix}.
        \label{eq:voltage_sensitivity}
    \end{split}
\end{equation}}%  
For a large value of $\omega_c$, $S_i$ in (\ref{eq:inner_outer_loop}) remains small over a broad frequency range and can be assumed to be negligible. Therefore, considering $\Delta \vv{v_c} = \vv{v_c} - \vv{v_0}$, we can rewrite (\ref{eq:inner_outer_loop}) as 
{\small\begin{align}
    \Delta \vv{\hat{v}_c} &=  
    K_{inv}\vv{\hat{i}_r}
    - S_v \vv{\hat{v}_0},
    &
    & \text{\normalsize where}
    &
    K_{inv} &= G_v S_v T_i.
    \label{eq:CC_Feedback}
\end{align}}%
\begin{remark}
$K_{inv}$ captures the closed-loop dynamics of the inverter as a compensator between $\vv{i_r}$ and $\Delta\vv{v_c}$ as shown in Fig. \ref{fig:Equivalent_Feedback}. $K_{inv}$ can represent any transfer function with a relative degree of at least two and the following form
{\small\begin{align}
    K_{inv} &=
    \frac{N(s)}{D(s)} =
    \frac{\omega_c}{C_i}
    \frac{(s + z_n)\cdots(s + z_0)}
    {(s + p_m)\cdots(s + p_0)},
    &
    n
    &\leq (m-2),
    \label{eq:controller_K_inv}
\end{align}}%
if voltage compensator $K_v = N_v(s)/D_v(s)$ is designed as
{\small\begin{align} 
    N_v(s) &= 
    D(s) - C_i s(s + \omega_c)N(s)
    /\omega_c,
    &
    D_v(s) &= N(s),
    \label{eq:controller_K_v_K_c}
\end{align}}%
and the inner closed-loop controller $K_i$ is set to  
{\small\begin{align}
    K_i &=
    \omega_c\frac{L_i s + R_i}{s},
    &
    &\text{\normalsize where}
    &
    \omega_c &= \sum_{i = 0}^m p_i - 
    \sum_{i = 0}^n z_i,
    \label{eq:Implement_K_i}
\end{align}}%
where $p_i$ and $z_i$ are poles and zeros of $K_{inv}$ in (\ref{eq:controller_K_inv}).
\end{remark}
We can cascade $K_{inv}$ with a series compensator $K_c$ to form the feedback compensator between $\vv{e}'$ and $\Delta\vv{v_c}$ shown in Fig. \ref{fig:Equivalent_Feedback}. In this case, comparing Fig. \ref{fig: Simple_Feedback} and Fig. \ref{fig:Equivalent_Feedback} show
{\small\begin{align}
    K_c K_{inv} = 
    \begin{bmatrix}
        K_c^d K_{inv}^d& 0 \\
        0 & K_c^q K_{inv}^q
    \end{bmatrix}
    =
    \begin{bmatrix}
        K^d & 0 \\
        0  & K_1^q
    \end{bmatrix}.
\end{align}}%
The equation above indicates that poles and zeros of the controllers $K^d$ and $K_1^q$ should be divided between $K_c$ and $K_{inv}$. We suggest that the inverter dynamics $K_{inv}$ only includes the smallest zero and three of the largest poles of $K^d$ and $K_1^q$ and $K_c$ include rest of poles and zeros. For example $K^d$ and $K_1^q$ in (\ref{eq:Sample_Controller}) can be divided into $K_{inv}$ and $K_c$ as follows
{\small\begin{align}
    K_{inv}^d &=
    \frac{\omega_c^d}{C_i}
    \frac{s + \alpha_v}{(s + \omega_d)^3},
    \quad
    K_{inv}^q =
   \frac{\omega_c^q}{C_i}
   \frac{s}{(s + \omega_2)(s + \omega_q)^2},
   \label{eq:Implement_K_inv}
    \\
    K_c^d &=
    \frac{C_i}{\omega_c^d}
    \left(
    \frac{2\sqrt{2}\omega_d^3}{\omega_m / Z}
    \right)
    \left(
    \frac{s + \omega_m}{s + 2\sqrt{2}(aZ)\beta_v}
    \right)
    \left(
    \frac{s + a\omega_d}{as + \omega_d}
    \right),
    \\
    K_c^q &=
    \frac{C_i}{\omega_c^q}
    \left(
    \frac{2\omega_q^2\sqrt{\omega_q^2 + \omega_2^2} }{\omega_m / Z}
    \right)
    \left(
    \frac{s + \omega_m}
    {s + \omega_1}
    \right)
    \left(
    \frac{s + a\omega_q}{as + \omega_q}
    \right),
\end{align}}%
where $\omega_c^d = 3\omega_d - \alpha_v$ and $\omega_c^q = 2\omega_q + \omega_2$. Finally, we implement $K_{inv}$ in (\ref{eq:Implement_K_inv}) by designing $K_v$ and $K_i$ as
{\small\begin{align}
    K_v^d &= C_i    
    \frac{(3\omega_d^2 + \alpha_v \omega_c^d)s
    + \omega_d^3}
    {\omega_c^d (s + \alpha_v)},
    &
    K_i^d = \omega_c^d \frac{L_i s + R_i}{s},
    \\
    K_v^q &=
    C_i
    \frac{(\omega_q^2 + 2\omega_2)s + \omega_2\omega_q^2}
    {\omega_c^q s},
    &
    K_i^q = \omega_c^q \frac{L_i s + R_i}{s},
\end{align}}
according to (\ref{eq:controller_K_v_K_c}) and (\ref{eq:Implement_K_i}).
\subsubsection{Generating the rotating angle}
Based on the Proposition \ref{prop:control_disturbance}, the control signal $u_\theta$ from $K_2^q$ encodes the $dq$ rotating angle $\theta$. Therefore, in practice, $u_\theta$ is implemented indirectly through $\theta$ as follows
{\small\begin{align}
    \theta(t) = \omega_0 t +
    \frac{1}{v_0} u_\theta(t)
    \pmod{2\pi}.
\end{align}}%
\subsection{Seamless Transition Between Operating Modes}
\begin{wrapfigure}{r}{0.5\linewidth}
    \includegraphics[width=1\linewidth]{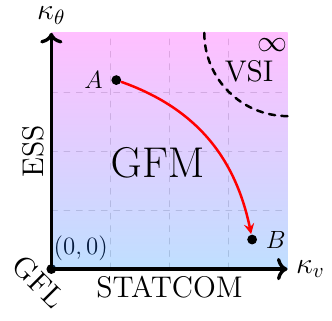}
    \caption{2d operating space parameterized by $(\kappa_v,\kappa_\theta)$.}
    \label{fig:Hybrid_Transition}
    \vspace{0em}
\end{wrapfigure}
We define $\kappa_v = \beta_v / \alpha_v$ and $\kappa_\theta = \beta_\theta / \alpha_\theta$, where ${\alpha_v, \beta_v}$ and ${\alpha_\theta, \beta_\theta}$ are the parameters in (\ref{eq:Sample_Controller}). The $(\kappa_v, \kappa_\theta)$ pair defines a 2D steady-state operating space, as shown in Fig. \ref{fig:Hybrid_Transition}. The y-axis ($\kappa_v = 0$) corresponds to an integrator in $K^d$, while the x-axis ($\kappa_\theta = 0$) corresponds to two integrators in $K^q$. According to Table \ref{tab:Mode_Integral}, the y-axis represents the ESS mode, the x-axis represents STATCOM operation, and the origin represents GFL mode. Moreover, each point within this 2D space specifies a unique combination of operating modes; for example, point A in Fig. \ref{fig:Hybrid_Transition} is closer to the ESS mode, while point B is closer to STATCOM operation.

Changing $\kappa = (\kappa_v,\kappa_\theta)$ allows seamless transitions between operating modes, as shown in Fig. \ref{fig:Hybrid_Transition}. The secondary control layer can leverage the mode transitions to create a flexible and adaptive framework that aligns with economic or dispatch goals in a dynamic environment. Although the controller in (\ref{eq:Sample_Controller}) ensures stability in each mode (pointwise stability), we must also ensure stability during mode transition, given a sufficiently slow change in the parameters $\kappa = (\kappa_v,\kappa_\theta)$ \cite{stilwell2002stability}.
\begin{propositionE}
\label{prop:mode_trans}
    Pointwise stability at each operating mode $\kappa = (\kappa_v,\kappa_\theta)$ is a sufficient condition to ensure the stability of the closed-loop system during mode transitions, provided that the parameters $\kappa = \left(\kappa_v, \kappa_\theta\right)$ change slowly enough.
\end{propositionE}
\begin{proofE}
    The seamless transition between operation modes involves the adjustment of controller parameters $K$, which results in a closed-loop linear parameter-varying (LPV) system. We guarantee the stability of the closed-loop, shown in Fig. \ref{fig: Simple_Feedback}, during mode transitions, limiting the rate of change of parameters. To begin, we rewrite the closed-loop system in (\ref{eq:GLS_Nonlinear}) in the following form:
    {\small\begin{align}
        \frac{d}{dt}x &= (A(\kappa) - B K_{min} C)x + B u_\psi,
        &
        y &= Cx
        \label{eq:transformed_FB_ss}
        \\
        u_\psi &= -\psi(t,y) + K_{min}y = -\psi_*(t,y)
        \label{eq:transformed_fb_nonlinear}
    \end{align}}%
    where $x = [x_g\; x_k ]^\top$ and $\{A,B,C\}$ triplet are defined as
    {\small\begin{align}
        A(\kappa) &=
        \begin{bmatrix}
            A_g + D_k & C_k \\
            -B_k & A_k(\kappa)
        \end{bmatrix},
        \;
        B = 
        \begin{bmatrix}
            I_2 \\ \mathbf{0}
        \end{bmatrix},
        \;
        C = 
        \begin{bmatrix}
            I_2 & \mathbf{0}
        \end{bmatrix}.
    \end{align}}%
    Moreover, the transformed feedback nonlinearity $\psi_*(\cdot,\cdot)$ in (\ref{eq:transformed_fb_nonlinear}) belongs to the sector $[0,K_*]$, where $K_* = K_{max} - K_{min} = 2 I_2 \Delta \omega_{max}$ (see (\ref{eq:sector_definition}) and (\ref{eq:freq_deviation})).\\
    The system in (\ref{eq:transformed_FB_ss}) has the following transfer matrix realization {\small\begin{equation}
        G_L S (I_2 + K_{min}G_L S)^{\text{-}1} 
        :=
        \left[
        \begin{array}{c|c}
        A_* & B \\
        \hline
        C & \mathbf{0}
        \end{array}
        \right].
    \end{equation}}%
    where $A_* = A(\kappa) - BK_{min}C$. If $Z(s)$ in (\ref{eq:Zs_SPR}) is SPR, then there exist a positive definite matrix $W$, matrix $L$ and a positive constant $0 < n$ such that $\{A_*,B,C\}$ satisfies
    {\small\begin{align}
        & W A_* + A_*^\top W = -L^\top L - n W ,
        \label{eq:Lyapunov_equation}
        \\
        & WB = C^\top K_* - \sqrt{2}L^\top.
        \label{eq:Lyapunov_SPR}
    \end{align}}
    We adopt the Lyapunov function $v(x) = x^TW(\kappa)x$ for the system in (\ref{eq:transformed_FB_ss}) where $W$ is the same as (\ref{eq:Lyapunov_equation}). The time derivative of $v(x)$ along the trajectories of $x(t)$ is
    {\small\begin{align}
        \Dot{v}(x) = x^T(\Dot{W}(\kappa) - L^\top L - n W)x -2x^\top WB \psi_*(t,y).
        \label{eq:lyapunov_derivative}
    \end{align}}%
    The nonlinearity $\psi_*(\cdot,\cdot)$ belongs to the sector $[0,K_*]$ therefore we add the positive term $-2\psi_*(t,y)^\top(\psi_*(t,y) - K_*y) > 0$ to the right-hand side of (\ref{eq:lyapunov_derivative}) to get the following inequality
    {\small\begin{align}
    \begin{split}
        \Dot{v}(x) &\leq 
        x^T\Dot{W}(\kappa) x - x^T L^\top L x
        - 2\psi_*^{\top}(t,y)\psi_*(t,y)
        \\
        & \hphantom{\leq}
        + 2x^\top(C^{\top}K_* - WB) \psi_*(t,y) - n v(x).
    \end{split}
    \end{align}}%
    Based on (\ref{eq:Lyapunov_SPR}), we complete the square term above to get
    {\small\begin{align}
        \Dot{v}(x) &\leq 
        x^T\Dot{W}(\kappa) x - x^\top M^\top M x - n v(x),
        \label{eq:Lyapunov_upper_bound}
    \end{align}}%
    where $Mx$ is defined below
    {\small\begin{align}
        Mx = Lx - \sqrt{2}\psi_*(t,y) = 
        \left(
        L - \sqrt{2}\Delta\omega
        \begin{bmatrix}
            0 & 1 \\
            -1 & 0
        \end{bmatrix}C
        \right)x.
    \end{align}}
    We apply Gr\"{o}nwall's inequality to demonstrate that the Lyapunov function $v(x)$ and the closed-loop trajectories remain bounded and stable during the transition. To achieve this, we need to bound $\Dot{v}(x)$ in terms of $v(x)$. This is accomplished by applying two consecutive bounds to (\ref{eq:Lyapunov_upper_bound}), as shown below
    {\small\begin{align}
        \Dot{v}(x) &\leq 
        x^T \Dot{W}(\kappa) x - n v(x) \leq q(t) v(x).
    \end{align}}%
    The term $q(t)$ is defined below by directly applying the Rayleigh-Ritz theorem to $x^T \Dot{W} x$.
    {\small\begin{equation}
        q(t) =  
        \frac{\rho(\Dot{W}(\kappa))}{\underline{\lambda}(W(\kappa))} - n.
        \label{eq:Gronwall_Bound}
    \end{equation}}%
    In above $\rho(\cdot)$ is the spectral radius, and $\underline{\lambda}(\cdot)$ indicates smallest eigenvalue.     Employing the Gr\"{o}nwall's inequality, we get the following bound on closed-loop trajectories
    {\small\begin{align}
        v(t,x) = x^\top W x &\leq 
        v(0,x_0)
        \exp{\left(\int_{0}^{t} q(\mu)d\mu\right)}.
    \end{align}}%
    Based on the above, the sufficient condition for Lyapunov stability is the convergence of the exponential term to zero. Therefore, $q(t) < 0$ provides a sufficient condition for stability. We convert this condition into an upper bound on the rate of change of $\{\kappa_v, \kappa_\theta\}$ through the following three steps: \\
    $\bullet$ In (\ref{eq:Gronwall_Bound}), we express $\Dot{W}$ as $\Dot{\kappa}_v \partial W/\partial \kappa_v + \Dot{\kappa}_\theta \partial W / \partial \kappa_\theta$, where the partial derivatives are applied element-wise.\\
    $\bullet$ The matrices $\partial W / \partial \kappa_v$ and $\partial W / \partial \kappa_\theta$ are symmetric, and for symmetric matrices, the spectral radius $\rho(\cdot)$ is equal to $\|\cdot\|_2$.\\
    $\bullet$ The matrix two-norm is subadditive.\\
    Based on (\ref{eq:Gronwall_Bound}) and the three steps outlined above, the following condition is sufficient to ensure $q(t) < 0$.
    {\small\begin{align}
    \left[ |\Dot{\kappa}_v|\; |\Dot{\kappa}_\theta| \right]
    \begin{bmatrix}
        \| \partial W(\kappa) / \partial \kappa_v \|_2 
        \\
        \| \partial W(\kappa) / \partial \kappa_\theta \|_2
    \end{bmatrix}
    \leq
    \underline{\lambda}(W(\kappa)) n.
    \end{align}}%
    For small $\{\Dot{\kappa}_v,\Dot{\kappa}_\theta\}$ we can satisfy the above inequality since the right-hand side is always defined and positive. $\blacksquare$
\end{proofE}
\begin{figure}
  \begin{center}
  \includegraphics[width=0.95\linewidth]{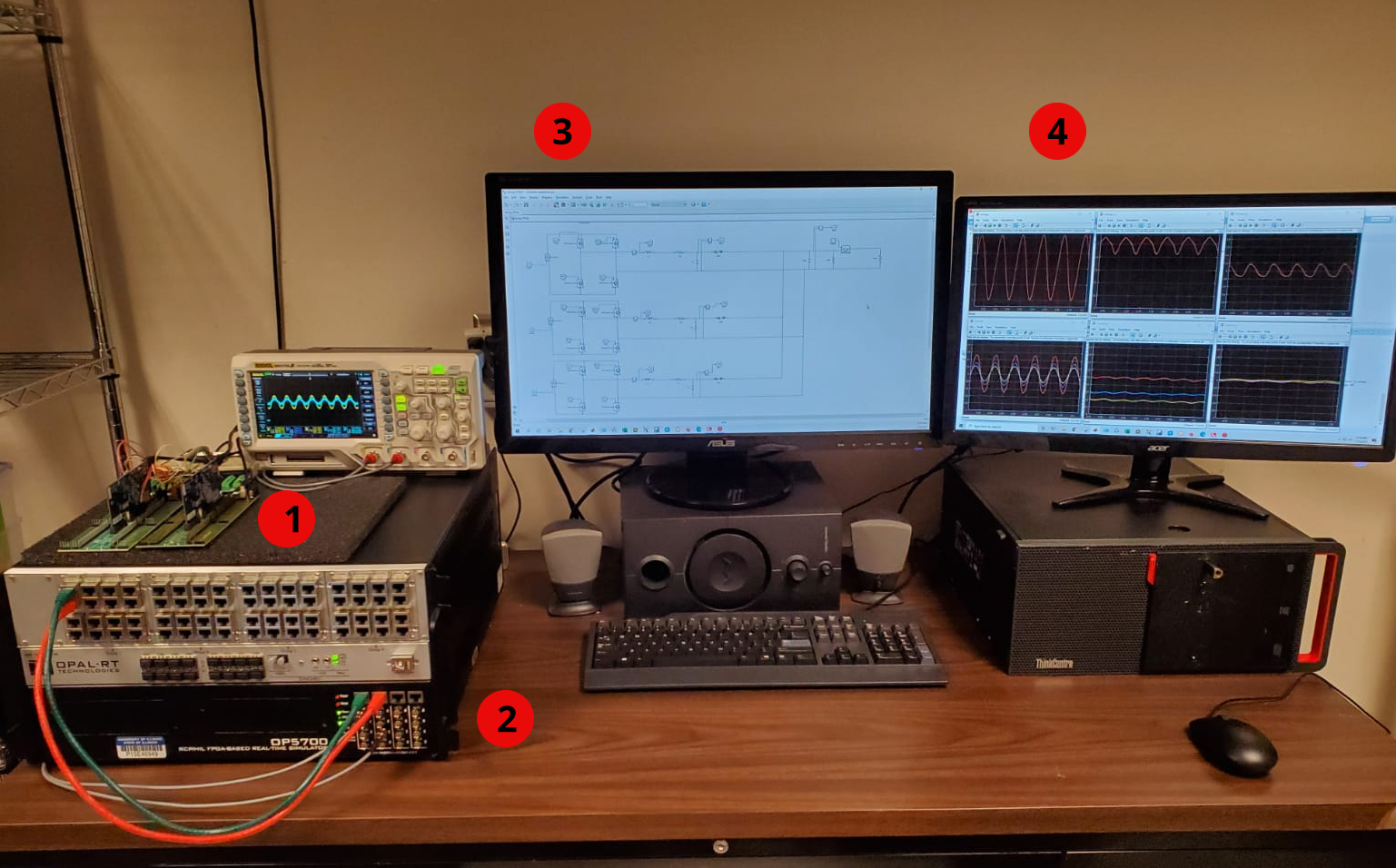}
    \caption{Experimental setup with (1) the controllers implemented on TMS320F28388D C2000 MCUs, (2) The OP5700 PHIL real-time simulation device, (3) switched model inverter and loads, (4) and real-time measurement signals.}
    \label{fig: Setup}
  \end{center}
  \vspace{-5mm}
\end{figure}
\begin{table}[ht]
    \vspace{0mm}
    \centering
    \caption{Simulation and Experimental Parameters}
    \begin{tabularx}{\columnwidth}{>{\centering\arraybackslash}m{0.15\columnwidth}
                                     | >{\centering\arraybackslash}m{0.05\columnwidth}
                                     | >{\centering\arraybackslash}m{0.07\columnwidth} 
                                     | >{\centering\arraybackslash}m{0.09\columnwidth}
                                     | >{\centering\arraybackslash}m{0.18\columnwidth}
                                     | >{\centering\arraybackslash}m{0.05\columnwidth} 
                                     | >{\centering\arraybackslash}m{0.08\columnwidth}}
    \toprule[1pt]
    \midrule[0.3pt]
        Parameter & Sym. & Sim. & Expt. & Parameter & Sym. & Value \\
    \midrule
        Filter capacitor & $C_i$ & 15$\mu$F & 15$\mu$F & Grid voltage (line-RMS) & $v_g$ & 120V \\
    \hline
        Filter inductor & $L_i$ & 1mH & 470$\mu$H & Fundamental Frequency & $\omega_g$ & 60Hz\\
    \hline
        Grid inductor & $L$ & 1mH & 470$\mu$H & Switching frequency & $f_{sw}$ & 50kHz \\
    \hline
        Grid resistance & $R$ & 1m$\Omega$ & 1.4m$\Omega$ & Sampling Frequency & $f_s$ & 50kHz \\
    \midrule[0.3pt]
    \bottomrule[1pt]
    \end{tabularx}
    \label{tab:Sim_Expt_Parameter}
    \vspace{-3mm}
\end{table}
\section{Simulation and Experimental Results}
We demonstrate the key features of our framework through OPAL-RT hardware-in-the-loop (HIL) real-time simulations shown in Fig. \ref{fig: Setup}, and experimental results. The experimental setup consists of three-phase inverters and All Control algorithms are implemented on the TI TMS320F28388D DSP.
\subsection{GFM On Grid Operation}
Operating GFM inverters in grid-tied mode improves the resilience of the power network to sudden islanding. However, accurate current tracking or sharing in grid-tied GFM depends on three key factors: grid condition (nominal or disturbed), line characteristics (inductive, complex, or resistive), and whether the line impedance is known. Here, a nominal grid means that there is no deviation in grid voltage and frequency, that is $\Delta v_g=\Delta\omega_g = 0$, as opposed to a disturbed grid condition. These factors create 24 unique scenarios for power regulation and power sharing, as detailed in Table \ref{tab: GFM_Power_SS}. 

In this section, we explore the capabilities of the proposed framework operating in grid-tied GFM mode, focusing on current tracking and sharing, inertial response, fault ride-through, and robustness.
\begin{table}[ht]
    \centering
    \begin{tabular}{cc|c|c|c|c|c|c|}
         \multicolumn{2}{c}{\multirow{2}{*}
         {\diagbox{Grid}{$Z \angle \phi_z$}}}
         & \multicolumn{2}{|c|}{Inductive} 
         & \multicolumn{2}{c|}{Complex} 
         & \multicolumn{2}{c|}{Resistive} \\
         && A & N/A & A & N/A & A & N/A \\
         \hline
         \multirow{2}{*}{Nominal}
         & $P$
         & \checkmark \checkmark & \checkmark \checkmark 
         & \checkmark \checkmark & \ding{56}
         & \checkmark \checkmark & \ding{56} \\
         \cline{2-8}
         & $Q$
         & \checkmark \checkmark & \ding{56} 
         & \checkmark \checkmark & \ding{56}
         & \checkmark \checkmark & \checkmark \checkmark \\
         \hline
         \multirow{2}{*}{Disturbed}
         & $P$
         & \checkmark & \checkmark 
         & \checkmark & \ding{56} 
         &  \checkmark & \ding{56} \\
         \cline{2-8}
         & $Q$
         & \checkmark & \ding{56} 
         & \ding{56} & \ding{56} 
         & \checkmark & \checkmark \\
    \end{tabular}
    \caption{Ability to achieve exact active ($P$) or reactive ($Q$) power regulation (\checkmark\checkmark) or power sharing (\checkmark) based on whether the exact value of line impedance is known (A) or unknown (N/A), under nominal and disturbed grid conditions.}
    \label{tab: GFM_Power_SS}
\end{table}
\subsubsection{Current tracking under nominal grid}
\begin{figure}[t]
    \centering
    \vspace{0mm}
    \includegraphics[width = 1\columnwidth]{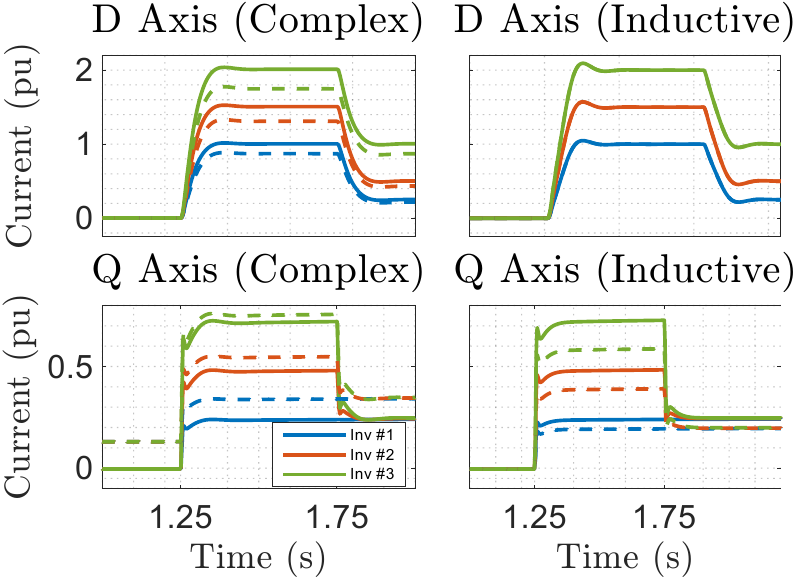}
    \caption{Exact current tracking of proposed GFM (solid) vs conventional GFM (dashed) for nominal grid operation.}
    \label{fig: Sharing_Complex}
    \vspace{0mm}
\end{figure}

Fig. \ref{fig: Sharing_Complex} shows a set of inverters with proposed GFM control (solid) that accurately tracks changing current setpoints for known inductive and complex line impedances. For the inductive line, knowledge of the impedance value is only required for exact regulation of the reactive current (see Table \ref{tab: GFM_Power_SS}).
\subsubsection{Current sharing under disturbed grid}
\begin{figure}[t]
    \centering
    \vspace{-4mm}
    \subfloat[]{
    \includegraphics[width = 0.48\columnwidth]{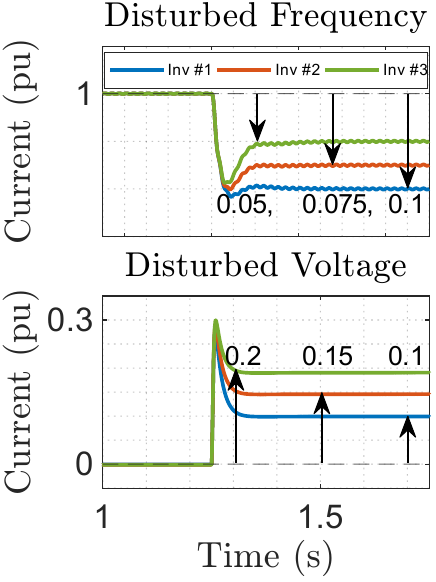}
    \label{fig:Power_Sharing_Deviation}
    }
    \subfloat[]{
    \includegraphics[width = 0.49\columnwidth]{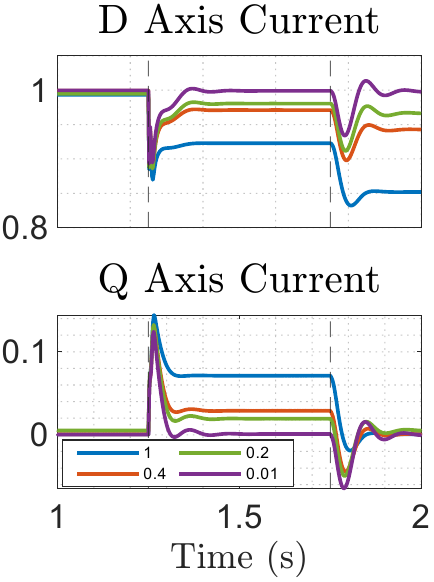}
    \label{fig:Improve_Sharing}
    }
    \caption{(a) Current droop under grid frequency and voltage disturbance. (b) Improved regulation for small $\kappa_v$ and $\kappa_\theta$.}
    \vspace{0mm}
\end{figure}

Fig. \ref{fig:Power_Sharing_Deviation} shows the inherent $P/f$, $Q/v$ droop behavior of the proposed framework for a predominantly inductive line impedance and when the voltage and frequency of the grid deviate by 0.1pu and 0.1 Hz, respectively.
\begin{figure}[!t]
    \centering
    \includegraphics[width = 1\columnwidth]{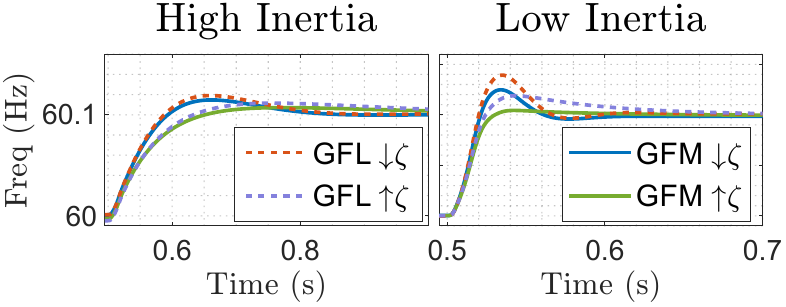}
    \caption{Inverter frequency response to a step change in the grid frequency for different values of $\omega_J$ and $\omega_J/\omega_1$.}
    \label{fig: Inverter_Freq_Response}
    \vspace{-5mm}
\end{figure}
In addition, the inverters in Fig. \ref{fig:Power_Sharing_Deviation} achieve power sharing ratios of $\gamma_1 = 0.22,\gamma_2 = 0.33,\gamma_1 = 0.45$ by designing $\{\alpha_v,\beta_v\}$ and $\{\alpha_\theta,\beta_\theta\}$ as specified in (\ref{eq:Kd_Droop}) and (\ref{eq:Kq_Droop}).
\begin{figure}[t]
    \centering
    \includegraphics[width = 0.95\columnwidth]{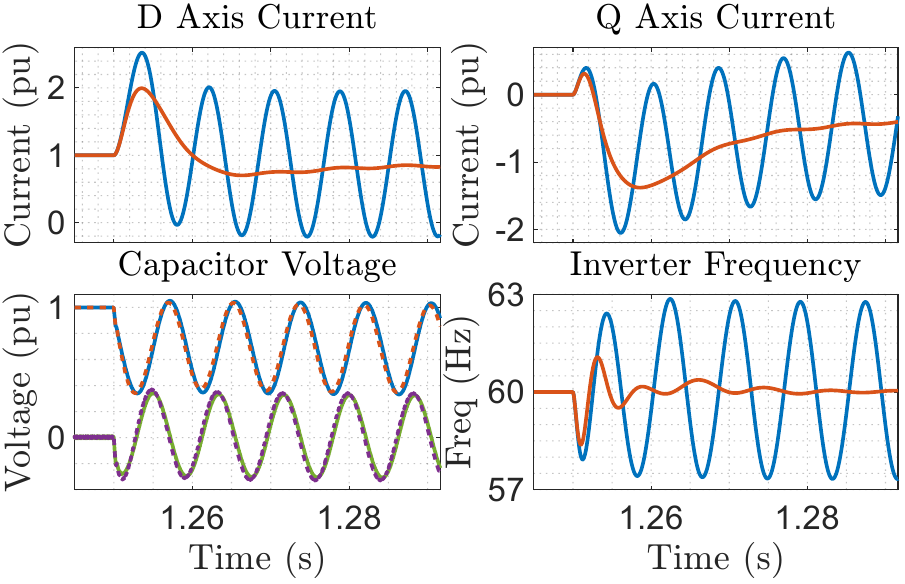}
    \caption{GFM Line-to-ground fault ride through by rejecting the 120 Hz ripple on inverter current and frequency.}
    \label{fig: L2G_Fault}
    \vspace{0mm}
\end{figure}
\begin{figure}[t]
    \centering
    \subfloat[]
    {
    \includegraphics[width = 0.95\columnwidth]{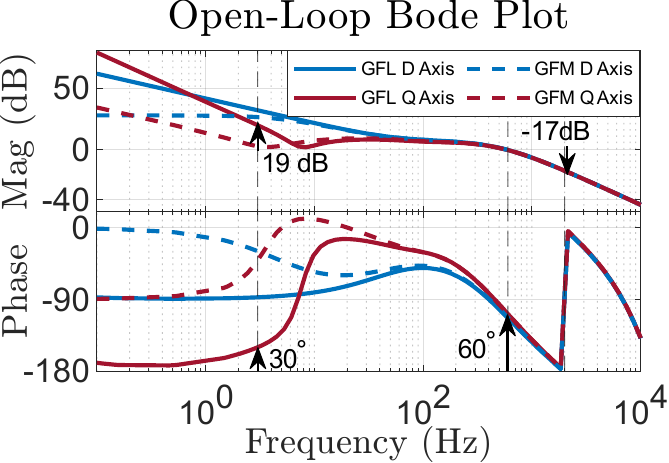}
    \label{fig:Robust_Weak_Grid_Bode}
    }
    \vspace{0em}
    \subfloat[]
    {
    \includegraphics[width = 0.48\columnwidth]{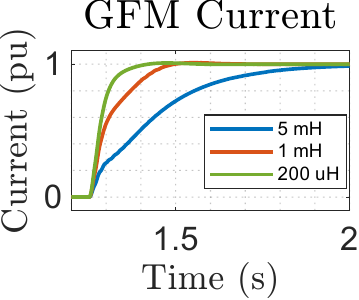}
    \label{fig:Robust_Weak_Grid_GFM}
    }
    \subfloat[]
    {
    \includegraphics[width = 0.48\columnwidth]{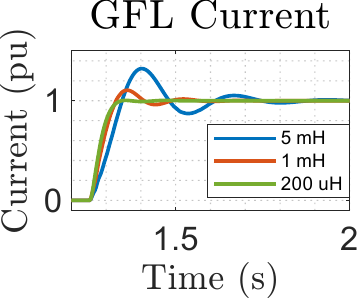}
    \label{fig:Robust_Weak_Grid_GFL}
    }
    \caption{(a) GFM and GFL $d$ and $q$ axes open-loops. (b) GFM and (c) GFL response to step-change in active current.}
    \label{fig:Robust_Bode}
    \vspace{0mm}
\end{figure}%
\begin{figure}[t]
    \centering
    \includegraphics[width = 0.9\columnwidth]{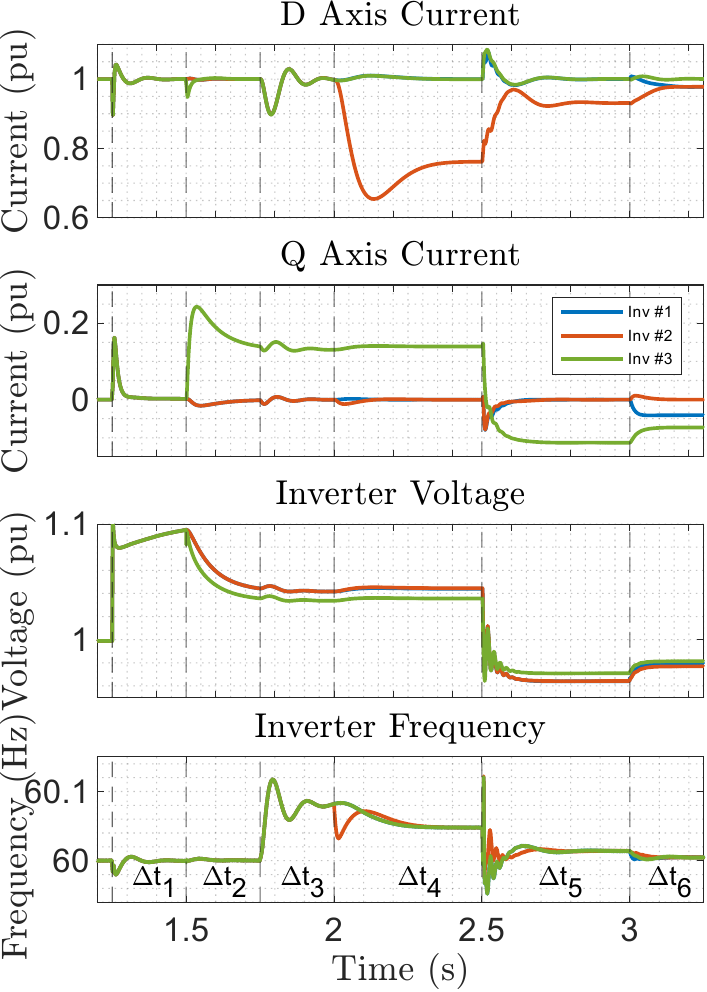}
    \caption{Seamless transition of set of inverters based on Table \ref{tab:Seamless_Trans_On_Grid} during on-grid operation.}  
    \label{fig: Seamless_Transition}
    \vspace{0mm}
\end{figure}
\subsubsection{Improving accuracy of current injection}
Reducing the sensitivity $\widetilde{S}$ at low frequencies improves current regulation, regardless of grid conditions or line impedance characteristics. This is achieved by decreasing $\{\kappa_v,\kappa_\theta\}$ to assert the GFL behavior of the inverter, as shown in Fig. \ref{fig:Hybrid_Transition}. Fig. \ref{fig:Improve_Sharing} shows the rapid improvement in current tracking during grid voltage disturbances at $t=1.25$ and frequency excursion at $t=1.75$ for different values of $\{\kappa_v,\kappa_\theta\}$ and complex line impedance.
\subsubsection{Inverter Inertial Frequency Response}
Based on Section \ref{sec:high_freq_correction}, we shape the frequency transient response of a GFL and GFM inverter for high inertia $\omega_J = 2\pi 5$ [rad/s], and low inertial $\omega_J = 2\pi 20$ [rad/s]. Fig. \ref{fig: Inverter_Freq_Response} shows the GFM and GFL step response to the grid frequency anomaly under different inertia and for both low damping $\downarrow\zeta \propto \omega_J/\omega_1 = 2$ and high damping $\uparrow\zeta \propto \omega_J/\omega_1 = 10$. For same inertia $\omega_J$ and damping $\omega_J/\omega_1$, GFM always exhibits a lower overshoot.
\subsubsection{Line-to-Ground Fault}
Line-to-ground fault adds second harmonic component to disturbance model in (\ref{eq:disturbance_type}) \cite{askarian2024enhanced}. GFM inverters, unlike GFL, lack the high bandwidth current control to handle current transients and ripple during fault. However, we can achieve fault ride-through by adding a second harmonic PR compensator into the controller. As discussed in Section \ref{sec: Extending_the_Steady_State_Analysis}, this makes $\widetilde{S}(j2\omega_0)$ small allowing the GFM to operate like a GFL at this particular frequency and reject the fault ripple. Fig. \ref{fig: L2G_Fault} shows that the proposed GFM inverter operate smoothly during a fault, rejecting the 120 Hz ripple on both the inverter output current and frequency. This is an improvement over conventional GFM, shown by the blue line in Fig. \ref{fig: L2G_Fault}.

\subsubsection{Robustness to Line Impedance Uncertainty}
The proposed framework is robust to line impedance uncertainty. Fig. \ref{fig:Robust_Weak_Grid_Bode}, shows that the $d$ and $q$ open-loops consistently achieve a $60^\circ$ phase margin and a $17$ dB gain margin in all modes. This allows the controller to tolerate a decrease in the magnitude of the line impedance, $Z$ in (\ref{eq:GM_Modified}), by a factor of $7$ ($-17$ dB), while increasing $Z$ does not cause instability. However, increasing line impedance reduces the $q$-axis open-loop gain, lowering the crossover frequency and phase margin in GFL mode, resulting in sluggish and oscillatory behavior. Fig. \ref{fig:Robust_Weak_Grid_Bode} shows that reducing the gain of the $q$ axis by $19$ dB results in a crossover frequency of 3 Hz and a phase margin of approximately $30^\circ$. Fig. \ref{fig:Robust_Weak_Grid_GFM} and \ref{fig:Robust_Weak_Grid_GFL} show a stable step response of the nominal controller (designed for $L_g = 1$ mH) for line impedances: $L_g = 5$ mH, $L_g = 1$ mH, and $L_g = 200$ $\mu$H. As expected, there is a noticeable decrease in the bandwidth and oscillation for higher line inductance in the GFL mode. 

\subsection{Inverter Seamless Mode Transition}
\subsubsection{On-Grid Transition and Ancillary Service}
A set of three GFL inverters transitions seamlessly and independently between the operating modes to provide ancillary services based on the grid conditions. Table \ref{tab:Seamless_Trans_On_Grid} details three main contingency events, voltage deviation $\{\Delta t_1,\Delta t_2\}$, frequency deviation $\{\Delta t_3,\Delta t_4\}$, and islanding $\{\Delta t_5,\Delta t_6\}$. As shown in Fig. \ref{fig: Seamless_Transition}, transitioning to STATCOM and ESS modes during $\Delta t_2$ and $\Delta t_4$ improves voltage and frequency regulation. During islanding, both the ESS and STATCOM intervene to stabilize the frequency and voltage of the formed islanded grid. Switching the last GFL to GFM further enhance the voltage and frequency stability of the islanded microgrid. Table \ref{tab:Seamless_Trans_On_Grid} details the statue of the inverters and the grid. Fig. \ref{fig: Seamless_Transition} illustrates a smooth transition and satisfactory transient response during key moments outlined in Table \ref{tab:Seamless_Trans_On_Grid}.
The proposed control design enables a collection of inverters to establish an islanded microgrid if at least one inverter is set to operate in GFM mode or if the group includes at least one STATCOM and one ESS.
\begin{table}[ht]
    \centering
    \begin{tabular}{c|c|c|c|c|c|c}
         & $\Delta t_1$ & $\Delta t_2$ 
         & $\Delta t_3$ & $\Delta t_4$
         & $\Delta t_5$ & $\Delta t_6$\\
         inv 1 & \cellcolor{blue!25} GFL & \cellcolor{blue!25} GFL 
         & \cellcolor{blue!25} GFL& \cellcolor{blue!25} GFL&
         \cellcolor{blue!25} GFL&\cellcolor{green!25} GFM\\
         inv 2 & \cellcolor{blue!25} GFL & \cellcolor{blue!25} GFL 
         & \cellcolor{blue!25} GFL&\cellcolor{yellow!25} \cellcolor{yellow!25} ESS&\cellcolor{yellow!25} ESS&\cellcolor{yellow!25} ESS\\
         inv 3 & \cellcolor{blue!25} GFL & \cellcolor{red!25} STAT & \cellcolor{red!25} STAT&\cellcolor{red!25} STAT&\cellcolor{red!25} STAT&\cellcolor{red!25} STAT\\
         $\Delta v_g$ &0.1 pu&0.1 pu&0.1 pu&0.1 pu
         &\multicolumn{2}{|c|}{\multirow{2}{*}{Off-Grid}} \\
         $\Delta\omega_g$ &0 Hz&0 Hz&0.1 Hz&0.1 Hz&
         \multicolumn{2}{|c|}{}
    \end{tabular}
    \caption{Statue of inverters and grid.}
    \label{tab:Seamless_Trans_On_Grid}
    \vspace{0mm}
\end{table}
\begin{figure}
    \centering
    \includegraphics[width = 0.9\columnwidth]{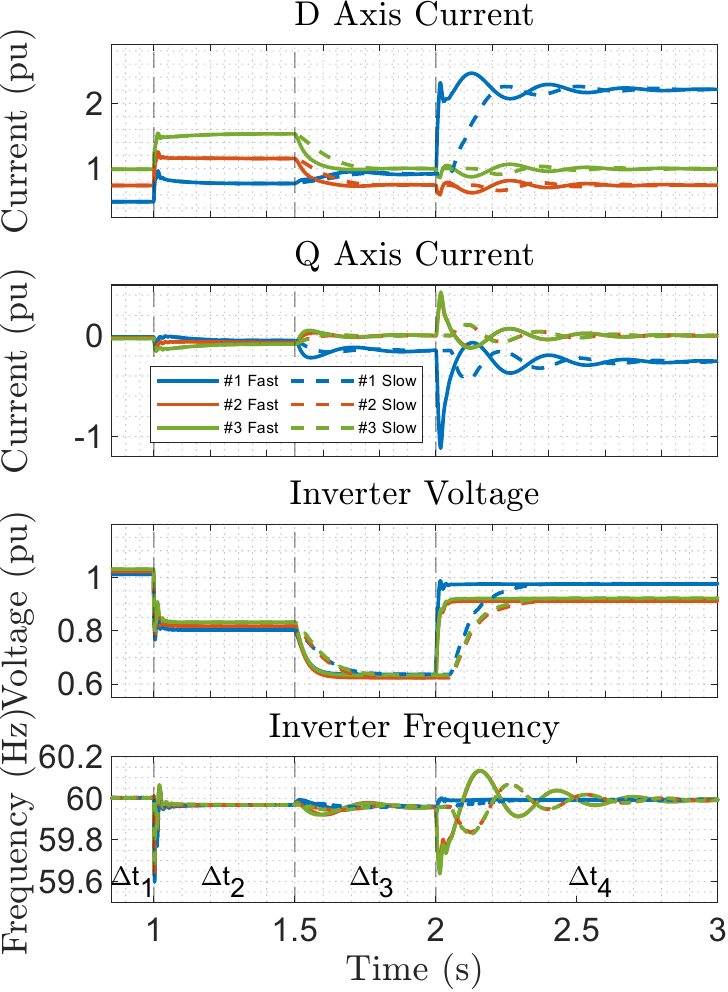}
    \caption{Seamless transition of a set of inverters based on Table \ref{tab:Seamless_Trans_Off_Grid} during off-grid operation.}
    \label{fig: Seamless_Transition_Off_Grid}
    \vspace{0mm}
\end{figure}
\subsubsection{Off-Grid Seamless Transition and Power Sharing}
A set of three GFM inverters form an islanded grid and maintain power sharing ratios ${\gamma_1 = 0.22, \gamma_2 = 0.33, \gamma_3 = 0.45}$ even when the load doubles during $\Delta t_2$. Subsequently, two of the inverters are forced to the GFL operation, which leads to a further drop in voltage. In the final step, to enhance the voltage and the frequency regulation we transition the remaining GFM to an ideal voltage source. Additionally, test results in Fig. \ref{fig: Seamless_Transition_Off_Grid} demonstrate that slower rate of mode transition results in reduced transients. Details of the inverters and the load are provided in Table \ref{tab:Seamless_Trans_On_Grid}.
\begin{table}[ht]
    \centering
    \vspace{0mm}
    \begin{tabular}{c|c|c|c|c|c|c|c}
         & $\gamma$ & $i_0^d$ & $i_0^q$ & $\Delta t_1$ & $\Delta t_2$ 
         & $\Delta t_3$ & $\Delta t_4$\\
         inv 1 & 0.22 & 0.5 pu & 0 pu & \cellcolor{green!25} GFM & \cellcolor{green!25} GFM 
         & \cellcolor{green!25} GFM& \cellcolor{orange!75} VSI\\
         inv 2 & 0.33 & 0.75 pu & 0 pu & \cellcolor{green!25} GFM & \cellcolor{green!25} GFM 
         & \cellcolor{blue!25} GFL&\cellcolor{yellow!25} \cellcolor{blue!25} GFL\\
         inv 3 & 0.45 & 1 pu& 0 pu & \cellcolor{green!25} GFM & \cellcolor{green!25} GFM & \cellcolor{blue!25} GFL&\cellcolor{blue!25} GFL \\
         \multicolumn{4}{c|}{Resistive Load} & 1 pu & 2 pu & 2 pu & 2 pu
    \end{tabular}
    \caption{Statue of inverters and load.}
    \label{tab:Seamless_Trans_Off_Grid}
    \vspace{0mm}
\end{table}

\section{Conclusion}
We introduced a novel and comprehensive control framework for inverters that enables seamless mode transitions and ensures stability, synchronization, and performance regardless of the operating mode. In our future research, we explore the integration of the proposed inverter control framework with an optimal secondary layer. Our main focus will be on developing a secondary layer that effectively manages the unpredictable and fluctuating patterns of energy generation and consumption by leveraging the mode transition at the inverter level.

\section{Appendix}

\printProofs

% References

\bibliographystyle{Bibliography/IEEEtranTIE}
\bibliography{Bibliography/IEEEabrv,Bibliography/BIB_xx-TIE-xxxx}\ %IEEEabrv instead of IEEEfull

\end{document}